\documentclass[12pt]{article} 
\usepackage{epsfig}
\usepackage{a4}
\usepackage{latexsym}
\usepackage{cite}

\usepackage{pslatex}
\usepackage[latin1]{inputenc}
\usepackage[T1]{fontenc}

\usepackage{color,colordvi}
\def\colour4colour#1{\Blue{#1}}

\renewcommand{\theequation}{\thesection.\arabic{equation}}
\newcommand{\gsim}{\raisebox{-0.07cm}{$\:\:\stackrel{>}{{\scriptstyle
 \sim}}\:\: $} }
\newcommand{\lsim}{\raisebox{-0.07cm}{$\:\:\stackrel{<}{{\scriptstyle
 \sim}}\:\: $} }
\newcommand{\beq}{\begin{equation}}
\newcommand{\eeq}{\end{equation}}
\newcommand{\bea}{\begin{eqnarray}}
\newcommand{\eea}{\end{eqnarray}}
\newcommand{\nn}{\nonumber}
\newcommand{\MSb}{$\overline{\mbox{MS}}$}
\newcommand{\as}{\alpha_{\rm s}}
\newcommand{\ar}{a_{\rm s}}
\newcommand{\ra}{\rightarrow}
\newcommand{\ep}{\epsilon}
\newcommand{\hspn}{{\hspace{-3mm}}}

\begin{document}
\setlength{\parskip}{0.25cm}
\setlength{\baselineskip}{0.54cm}

\def\Fone{{F_{\:\! 1}}}
\def\Ftwo{{F_{\:\! 2}}}
\def\FL{{F_{\:\! L}}}
\def\F3{{F_{\:\! 3}}}
\def\Qs{{Q^{\, 2}}}
\def\GeV2{{\mbox{GeV}^{\:\!2}}}
\def\DDk{{{\cal D}_{\:\!k}}}
\def\DD#1{{{\cal D}_{\:\! #1}}}
\def\z#1{{\zeta_{\:\! #1}}}
\def\zss{{\zeta_{2}^{\,2}}}
\def\zst{{\zeta_{3}^{\,2}}}
\def\zts{{\zeta_{2}^{\,3}}}
\def\ca{{C^{}_A}}
\def\cas{{C^{\, 2}_A}}
\def\cat{{C^{\, 3}_A}}
\def\cf{{C^{}_F}}
\def\cfs{{C^{\, 2}_F}}
\def\cft{{C^{\, 3}_F}}
\def\nf{{n^{}_{\! f}}}
\def\nfs{{n^{\,2}_{\! f}}}
\def\nft{{n^{\,3}_{\! f}}}
\def\dabc2{{d^{\:\!abc}d_{abc}}}
\def\dabcnc{{{d^{\:\!abc}d_{abc}}\over{n_c}}}
\def\fl11{fl_{11}}
\def\fl02{fl_{02}}

\def\S(#1){{{S}_{#1}}}
\def\Ss(#1,#2){{{S}_{#1,#2}}}
\def\Sss(#1,#2,#3){{{S}_{#1,#2,#3}}}
\def\Ssss(#1,#2,#3,#4){{{S}_{#1,#2,#3,#4}}}
\def\pqq(#1){p_{\rm{qq}}(#1)}
\def\gfunct#1{{g}_{#1}^{}}
\def\H(#1){{\rm{H}}_{#1}}
\def\Hh(#1,#2){{\rm{H}}_{#1,#2}}
\def\Hhh(#1,#2,#3){{\rm{H}}_{#1,#2,#3}}
\def\Hhhh(#1,#2,#3,#4){{\rm{H}}_{#1,#2,#3,#4}}
\def\Hhhhh(#1,#2,#3,#4,#5){{\rm{H}}_{#1,#2,#3,#4,#5}}

\def\gqqz{\gamma_{\,\rm qq}^{\,(0)}}
\def\gnso{\gamma_{\,\rm v}^{\,(1)}}
\def\gnst{\gamma_{\,\rm v}^{\,(2)}}
\def\ctqo{c_{3,\pm}^{(1)}}
\def\ctnt{c_{3,-}^{(2)}}
\def\ctnd{c_{3,-}^{(3)}}
\def\atnt{a_{3,-}^{(2)}}
\def\btqo{b_{3,\pm}^{(1)}}
\def\atqo{a_{3,\pm}^{(1)}}

\begin{titlepage}
\noindent
DESY 08-197, SFB/CPP-08-105 \hfill {\tt arXiv:0812.4168 [hep-ph]}\\
NIKHEF 08-032,~ LTH 815 \\
December 2008 \\
\vspace{1.3cm}
\begin{center}
\LARGE
{\bf Third-order QCD corrections to the}\\
\vspace{0.15cm}
{\bf charged-current structure function $F_3$} \\
\vspace{1.7cm}
\large
S. Moch$^{\, a}$, J.A.M. Vermaseren$^{\, b}$ and A. Vogt$^{\, c}$\\
\vspace{1.4cm}
\normalsize
{\it $^a$Deutsches Elektronensynchrotron DESY \\
\vspace{0.1cm}
Platanenallee 6, D--15738 Zeuthen, Germany}\\
\vspace{0.5cm}
{\it $^b$NIKHEF Theory Group \\
\vspace{0.1cm}
Kruislaan 409, 1098 SJ Amsterdam, The Netherlands} \\
\vspace{0.5cm}
{\it $^c$Department of Mathematical Sciences, University of Liverpool \\
\vspace{0.1cm}
Liverpool L69 3BX, United Kingdom}\\[1.2cm]
\vfill
\large
{\bf Abstract}
\vspace{-0.2cm}
\end{center}
We compute the coefficient function for the charge-averaged $W^\pm$-exchange
structure function $\F3$ in deep-inelastic scattering (DIS) to the third order 
in massless perturbative QCD.  Our new three-loop contribution to this quantity
forms, at not too small values of the Bjorken variable~$x$, the dominant part 
of the next-to-next-to-next-to-leading order corrections. It thus facilitates 
improved determinations of the strong coupling $\as$ and of $1/\Qs$ power 
corrections from scaling violations measured in neutrino-nucleon DIS.  The 
expansion of $\F3$ in powers of $\as$ is stable at all values of $x$ relevant 
to measurements at high scales $\Qs$.  At small $x$ the third-order coefficient
function is dominated by diagrams with the colour structure $\:\!\dabc2\:\!$ 
not present at lower orders. At large $x$ the coefficient function for $\F3$ is
identical to that of $\Fone$ up to terms vanishing for $x \ra 1$.
   
\vfill
\end{titlepage}
%
%
\setcounter{equation}{0}
\section{Introduction}
\label{sec:intro}
%
%
Structure functions in inclusive deep-inelastic lepton-nucleon scattering (DIS)
are among the most important observables probing QCD, the theory of the strong
interaction. Present data on these quantities can provide accurate information 
about the quark and gluon momentum distributions in the proton down momentum
fractions $x \approx 10^{-4}$, see Ref.~\cite{PDG2008}. These distributions, in 
turn, are indispensable ingredients for the analysis of all hard (high scale%
$/$mass) scattering processes at proton--(anti-)$\:\!$proton colliders, cf.~%
Refs.~\cite{Campbell:2006wx,Moch:2008dt}, such as {\sc Tevatron} and the LHC 
which will form the high-energy frontier of particle physics for at least the 
next fifteen years. Structure functions are also very well suited for precision
determinations of the strong coupling constant $\as$, one of the fundamental 
parameters of our description of nature, in the framework of perturbative QCD.

Inclusive quantities such as structure functions are the observables best 
accessible to field-theoretic calculations to `high orders' (today: two loops
and beyond) in the expansion in powers of the coupling constant. Indeed, the
second-order partonic cross sections (coefficient functions) for inclusive DIS
were completed as early as 1991/2 \cite{vanNeerven:1991nn,Zijlstra:1991qc,%
Zijlstra:1992kj}, while corresponding quantities for jet shapes in $e^+e^-$ 
collisions have been presented only very recently 
\cite{GehrmannDeRidder:2007bj,GehrmannDeRidder:2007hr,Weinzierl:2008iv}.
At large $x$ (with hindsight:\linebreak $x > 10^{-2}$) the former quantities 
are the dominant part of the next-to-next-to-leading order (NNLO) contributions 
in renormalization-group improved perturbation theory. They are not sufficient, 
though, for NNLO analyses over the full $x$-range opened up by HERA and 
required for the LHC.

The three-loop corrections for inclusive DIS -- required to complete the NNLO
framework and to provide the dominant next-to-next-to-next-to-leading order
(N$^3$LO) corrections at large-$x$ enabling a perturbative accuracy of 1\% for
$\as$ determinations -- have been the subject of a long-term research program,
which started from sum rules \cite{Larin:1991zw,Larin:1991tj}, and proceeded 
via low integer Mellin moments \cite{Larin:1994vu,Larin:1997wd,Retey:2000nq} to
the computation of the exact expressions for all NNLO splitting functions 
\cite{Moch:2004pa,Vogt:2004mw} and the third-order coefficient functions for 
the structure functions $\FL$ and $\Ftwo$ \cite{Moch:2004xu,Vermaseren:2005qc}.
In the present article, we extend the latter results to the most important 
structure function not covered by Ref.~\cite{Vermaseren:2005qc}, the 
vector$\,$--$\,$axial-vector interference structure function $\F3$ in 
charged-current, specifically ($W^+\! + W^-$)-exchange DIS measured with high 
precision in neutrino-nucleon DIS \cite{PDG2008}.

The remainder of this article is organized as follows: In Section~2 we briefly
recall the general formalism for the calculation of the coefficient function
for $\F3$ and discuss some aspects specific to the present computation. We 
then write down the coefficient functions for $F_3^{\,W^++W^-\!}$ in Section~3. 
The second- and third-order quantities are presented via compact and accurate
parametrizations. We also address the behaviour of the third-order coefficient 
functions at large and small $x$, stressing the importance of the $\dabc2$ 
contribution not present at lower orders. This and other numerical effects are 
then illustrated in Section~4 where we assess the size of the higher-order 
corrections. Finally we summarize our results in Section~5 and close with a 
brief outlook on possible future improvements on the present accuracy. 
Appendix~A contains the very lengthy exact expression of our new third-order 
coefficient function, and Appendix~B complements the discussion of $\F3$ in 
Section 3 by providing the subleading large-$x$ logarithms for the case of 
$\Ftwo$. It turns out that the quark coefficient functions for $\F3$ and 
$2x\Fone \equiv \Ftwo - \FL$ in charged-current DIS are the same in the 
large-$x$ limit, specifically that $C_3(x) = C_1(x) + {\cal O}(1-x)\,$ holds to
(at least) order $\as^{\,3}$.
%
%
\setcounter{equation}{0}
\section{Formalism and calculation}
\label{sec:general}
%
%
We are interested in unpolarized charged-current deep-inelastic scattering 
(DIS), i.e., the reaction
\bea
\label{dis}
  l(k) \:+\: {\rm nucl}(p) \:\:\ra\:\: l^{\,\prime} (k^{\,\prime}) \:+\: X 
  \:\: ,
\eea
where $l$, $l^{\,\prime}$ and `nucl' denote a charged lepton, its (anti-)$\,$%
neutrino (in this or the opposite order) and a nucleon with respective momenta 
$k$, $k^{\,\prime}$ and $p$. $X$ stands for all hadronic states allowed by 
quantum number conservation.
The inclusive cross section for the process~(\ref{dis}) can be written as
$ d \sigma \:\sim\: L^{\,\mu\nu} W_{\mu\nu\,}$ in terms of leptonic and
hadronic tensors $L_{\mu\nu}$ and $W_{\mu\nu}$. The former tensor is well 
documented in the literature for both pure electromagnetic and electroweak 
gauge-boson exchange, see, e.g., Ref.~\cite{PDG2008}.
Thus there is no need to consider it here.

Our focus is on the (spin-averaged) hadronic tensor $W_{\mu\nu}$ which can be
expanded to define the structure functions $F_{\,2,\:3\:,L}$, viz
\bea
\label{htensor}
  W_{\mu\nu}(p,q) & = & \frac{1}{4\pi}
    \int \! d^{\,4}z\; {\rm{e}}^{\,{\rm{i}}q \cdot z} \,
    \langle {\,\rm{nucl,}\,p}\vert J_{\mu}^\dagger(z) J_{\nu}(0)\vert \,
    {\rm{nucl,}\,p}\rangle  \nn \\[1mm]
  & = &
    e_{\mu\nu}\, \frac{1}{2x}\: F_{L}(x,\Qs) \:+\:
    d_{\mu\nu}\, \frac{1}{2x}\: F_{2}(x,\Qs) \:+\: 
   {\rm{i}} \,\ep_{\mu\nu\alpha\beta}\: \frac{p^\alpha q^\beta}{p\cdot q}\: 
    F_{3}(x,\Qs) \:\: .
\eea
Here  $q$ denotes the momentum transferred by the $W$-boson, with $\Qs \equiv
-q^2 > 0$, and $J_{\mu}$ represents the weak current. The Bjorken variable is 
defined as $x=\Qs/ (2\,p\cdot q)$ with $0 < x \leq 1$. We do not consider the 
functions $F_{\,2,\:L}$ corresponding to the symmetric $e_{\mu\nu}$
and $d_{\mu\nu}$ in this article, see Ref.~\cite{Vermaseren:2005qc}. Instead we
address the structure function $\F3$ associated to the totally antisymmetric 
tensor $\ep_{\mu\nu\alpha\beta}$ which arises from the vector$/$axial-vector 
interference of the two $V\!-\!A$ currents. Specifically we are interested here
in the charged-averaged quantity $F_3^{\,W^+ + W^-}$ which has been accurately
measured in neutrino-nucleon DIS off (almost) isoscalar targets \cite{PDG2008}
(recall that $F_3^{\,W^+ - W^-}$ is small in this case, being proportional to
a flavour asymmetry in the quark sea).

As in the previous calculations in Refs.~\cite{Bardeen:1978yd,Moch:1999eb,%
Moch:2004pa,Vogt:2004mw,Moch:2004xu,Vermaseren:2005qc}, we derive analytic 
expressions for the Mellin moments of the perturbative contribution to  
structure functions, which for the present case are defined as
\beq
\label{F3mom}
  F_{3}^{\,W^++W^-}(N,\Qs) \:\;=\:\; \int_0^1 \! dx\; x^{\,N-1}\, 
  F_{3}^{\,W^++W^-}(x,\Qs)\:\: .
\eeq
To this end we make use of the optical theorem relating the hadronic tensor in 
Eq.~(\ref{htensor}) to the imaginary part of the forward Compton amplitude 
$T_{\mu\nu}$ of virtual gauge-boson$\,$--$\,$nucleon scattering. This amplitude
is expressed in terms of a time-ordered product of two local currents to 
which standard perturbation theory applies,
\beq
\label{fcompton}
  T_{\mu\nu}(p,q) \:\;=\:\; {\rm{i}} \int d^4z\: {\rm{e}}^{\,{\rm{i}}q \cdot z}
\langle {\,{\rm nucl},\,p} \vert\:T\! \left( J^{\dagger}_{\mu}(z)J_{\nu}(0) 
 \right) \vert {{\rm nucl},\,p}\rangle \:\: .
\eeq
In $D=4-2\ep$ dimensions 
the structure function $\F3$ is obtained from $T_{\mu\nu}$ using the projection
\begin{eqnarray}
\label{proj3}
 P_{3}^{\,\mu\nu} &\! = \!& - {\rm{i}}\: \frac{1}{(1-2\ep)(1-\ep)}
 \; \varepsilon^{\,\mu\nu\alpha\beta}\: \frac{p_\alpha q_\beta}{2\,p\cdot q}
 \:\: .
\end{eqnarray}
As discussed in detail in the literature, see, e.g., 
Refs.~\cite{Bardeen:1978yd,Moch:1999eb}, the operator-product expansion (OPE) 
can be applied to the product of currents in Eq.~(\ref{fcompton}) together with
a dispersion relation. For the structure function considered here this 
procedure finally yields
\beq
\label{F3mellin}
  \frac{1 - (-1)^N}{2}\: F_{3}^{\,W^++W^-}(N,\Qs) \:\: = \:\:
  C_3^{\,-} \left(N, \frac{\Qs}{\mu^{\,2}}, \as \right) A_{\rm v,\, nucl}
  (N,\mu^2) \:\: . 
\eeq
One thus obtains the odd-$N$ moments (\ref{F3mom}) which uniquely fix all 
moments, and hence the complete $x$-dependence, by analytic continuation. The 
notation $C_3^{\,-}$ for the coefficient function of $F_{3}^{\,W^++W^-}$ 
reflects this basis of odd moments. $A_{\rm v,\, nucl}(N,\mu^2)$ in 
Eq.~(\ref{F3mellin}) is the nucleon matrix elements of the combination 
$O_{\rm v}$ of local spin-$N$ twist-two quark operators corresponding to the 
total valence quark distribution, renormalized at the scale $\mu$. $1/\Qs$ 
corrections have been disregarded in Eq.~(\ref{F3mellin}).

As usual, we perform the renormalization in the modified~\cite{Bardeen:1978yd}
minimal subtraction scheme \cite{'tHooft:1973mm} in dimensional regularization.
Hence also the renormalized coupling $\as$ in Eq.(\ref{F3mellin}) refers to 
$D=4-2\ep\,$ dimensions, i.e., its scale dependence is given by
\beq
\label{arun}
  \frac{d}{d \ln \mu^{\,2}}\: \frac{\as}{4\pi} \:\: \equiv \:\:
  \frac{d\,\ar}{d \ln \mu^{\,2}} \:\: = \:\: \mbox{} - \epsilon\, \ar
  - \beta_0\, \ar^{\,2} - \beta_1\, \ar^{\,3} - \beta_2\, \ar^{\,4} 
  - \beta_3\, \ar^{\,5} - \ldots 
\eeq
where $\beta_{n}$ denote the coefficients of the usual four-dimensional \MSb\ 
beta function of QCD. The renormalization to the third order requires the 
coefficients up to $\beta_2$~\cite{Tarasov:1980au,Larin:1993tp}, and the
consistent application of the resulting three-loop (N$^3$LO) coefficient 
functions in data analyses even $\beta_3$, the highest contribution to 
Eq.~(\ref{arun}) computed so far \cite{vanRitbergen:1997va,Czakon:2004bu}.

There is one delicate issue entering the operator renormalization which 
deserves special attention. This is the presence of $\gamma_{\:\!5}$ due to 
the axial-vector coupling of the $W$-boson. As before in Refs.\ 
\cite{Larin:1991tj,Moch:1999eb,Retey:2000nq,Moch:2004pa} we employ the 
so-called Larin scheme \cite{Larin:1991tj,Larin:1993tq} which, at the present 
level, is equivalent to, but computationally more convenient than the original 
prescription~\cite{g5-HVBM1,g5-HVBM2} for consistently dealing with 
$\gamma_{\:\!5}$ in dimensional regularization. See also the discussion in 
Ref.~\cite{Zijlstra:1992kj}.
Therefore we need to perform a special renormalization, based on relating 
vector and axial-vector currents, in order to restore the axial Ward-identities,
\beq
\label{OA5ren}
  O_{\rm v} \:\: \ra \:\: Z_{5}\, Z_{A}\, O_{\rm v} \:\: .
\eeq
The constant $Z_A$ for the axial renormalization and the finite 
renormalization $Z_5$ due to this treatment of $\gamma_{\:\!5}$ in the 
\MSb-scheme are known to three loops~\cite{Larin:1991tj,Larin:1993tq}. The 
explicit expressions read
\bea
\label{ZA}
  \!Z_{A} &\!=\!& 
  1
  + a_{\rm s}^2 \* {1 \over \epsilon} \* \bigg[ \:
            {22 \over 3}\:\*\ca\*\cf 
          - {4 \over 3}\:\*\cf\*\nf 
    \bigg]
  + a_{\rm s}^3 \* \bigg[ \:
    {1 \over \epsilon^2} \* 
    \bigg(
          - {484 \over 27}\: \* \cf \* \cas 
          + {176 \over 27}\: \* \cf \* \ca \* \nf
          - {16 \over 27}\: \* \cf \* \nfs
    \bigg)
\nonumber \\[1mm] & & \mbox{} 
    + {1 \over \epsilon} \* 
    \bigg(
          - {308 \over 9}\: \* \cfs \* \ca
          +  {3578 \over 81}\: \* \cf \* \cas 
          + {32 \over 9}\: \* \cfs \* \nf
          - {832 \over 81}\: \* \cf \* \ca \* \nf
          + {8 \over 81}\: \* \cf \* \nfs
    \bigg)
    \bigg]
 \:\: , \\[3mm] 
\label{Z5}
  \!Z_{5} &\!=\!& 
  1
  + a_{\rm s} \* \bigg[
          - 4\, \* \cf 
          - 10\, \* \epsilon \* \cf 
          + \epsilon^2 \* \cf \* (
          - 22 
          + 2\, \* \z2) 
  \bigg]
  + a_{\rm s}^2 \* \bigg[ 
            22\, \* \cfs
          - {107 \over 9}\: \* \cf \* \ca
          + {2 \over 9}\: \* \cf \* \nf
\nonumber \\[1mm] & & \mbox{} 
       + \epsilon  \*  \bigg(
            \cfs\, \* (132 
          - 48\, \* \z3)
          + \cf \* \ca \* \bigg(
          - {7229 \over 54} 
          + 48\, \* \z3 \bigg)
          + {331 \over 27}\: \* \cf \* \nf
          \bigg)
  \bigg]
\nonumber \\[1mm] & & \mbox{} 
  + a_{\rm s}^3 \* \bigg[
            \cft \* \bigg(
          - {370 \over 3} 
          + 96\, \* \z3 \bigg)
          + \cfs \* \ca \* \bigg( \:
            {5834 \over 27} 
          - 160\, \* \z3 \bigg)
          + \cf \* \cas \* \bigg(
          - {2147 \over 27} 
          + 56\, \* \z3 \bigg)
\nonumber \\[1mm] & & \mbox{} 
          + \cfs \* \nf \* \bigg(
          - {62 \over 27} 
          - {32 \over 3}\: \* \z3 \bigg)
          + \ca \* \cf \* \nf \* \bigg( \:
            {356 \over 81} 
          + {32 \over 3}\: \* \z3 \bigg)
          + {52 \over 81}\: \* \cf \* \nfs \,
  \bigg]
  \:\: .
\end{eqnarray}
Both these expansions are given in terms of the renormalized coupling 
(\ref{arun}), and $Z_5$ is expressed to exactly the depth in $\ep$ implemented 
in the present calculation (see below Eq.~(\ref{T3-3}) for a comment on the 
positive powers of $\ep$ absent in Ref.~\cite{Larin:1991tj}$\,$).

We are now ready to briefly address the standard part of the renormalization of
the spin-$N$ operators $O_{\rm v}$ entering Eq.~(\ref{F3mellin}),  
\beq
\label{Oren}
  O_{\rm v}^{\,\rm ren} \:\: = \:\: Z_{\rm v}\: O_{\rm v}^{\,\rm bare} 
  \, ,
\eeq
where (as in some equations below) we have suppressed the dependence on $N$, 
and exploited the fact that $O_{\rm v}$ does not mix with other operators under
renomalization. The corresponding (non-singlet) anomalous dimensions 
$\gamma_{\,\rm v}$, governing the scale dependence of the operators $O_{\rm v}$
via
\beq
\label{gamma_v}
  \frac{d}{d \ln \mu^{\,2} }\: O_{\rm v}^{\,\rm ren } \:\: = \:\: 
  - \,\gamma_{\,\rm v}\, O_{\rm v}^{\,\rm ren } \:\: ,
\eeq
are connected to these renormalization constants in the standard way,
\beq
  \label{gamZv}
  \gamma_{\rm v} \:\: = \:\: -\,\bigg(\: \frac{d }{d\ln\mu^{\,2} }\: Z_{\rm v} 
  \bigg)\: Z^{-1}_{\rm v} \:\: = \:\: \sum_{l=0}^{\infty}\,
  a_{s}^{\,l+1}\, \gamma_{\rm v}^{\,(l)} \:\: .
\eeq
The reader is referred to Ref.~\cite{Moch:2004pa} for the relation of the 
expansion coefficients $\gamma_{\rm v}^{\,(l)}$ to those of other non-singlet 
combinations. In terms of these quantities, the \MSb\ renormalization factor 
$Z_{\rm v}$ in Eq.~(\ref{Oren}) is given by the Laurent series
\bea
\label{Zv3}
  Z_{\rm v} & = &
    1 \: + \: \:\ar\, \frac{1}{\ep}\,\gamma_{\,\rm v}^{\,(0)}
    \: + \: a_{\rm s}^2 \,\left[\, \frac{1}{2\ep^2}\,
    \left\{ \left(\gamma_{\,\rm v}^{\,(0)} - \beta_0 \right)
    \gamma_{\,\rm v}^{\,(0)} \right\}
    + \frac{1}{2\ep}\, \gamma_{\,\rm v}^{\,(1)} \right]
  \nn \\[1mm] & & \mbox{} + \:
  a_{\rm s}^3 \,\left[\, \frac{1}{6\ep^3}\,
    \left\{ \left( \gamma_{\,\rm v}^{\,(0)} - 2 \beta_0 \right)
    \left( \gamma_{\,\rm v}^{\,(0)} - \beta_0 \right)
    \gamma_{\,\rm v}^{\,(0)} \right\} \right.
  \nn \\[1mm] & & \left. \mbox{} \quad\quad \! + \:
  \frac{1}{6\ep^2}\, \left\{ 3\, \gamma_{\,\rm v}^{\,(0)}
    \gamma_{\,\rm v}^{\,(1)} - 2 \beta_0\, \gamma_{\,\rm v}^{\,(1)}
    - 2 \beta_1\, \gamma_{\,\rm v}^{\,(0)} \right\} \: + \:
    \frac{1}{3\ep}\, \gamma_{\,\rm v}^{\,(2)} \right] \:\: .
\eea

Of course, the perturbative calculation of the anomalous dimension 
(\ref{gamma_v}) and the coefficient function $C_3^{\,-}$  cannot proceed via
Eq.~(\ref{F3mellin}) including the non-perturbative nucleon states 
$\vert {\rm{nucl,}\,p}\rangle$. However, as the OPE represents an operator 
relation, the calculation can be performed using quark states instead. Hence
we apply the Lorentz projector~(\ref{proj3}) to the forward virtual-$W$--quark
Compton amplitude ${\cal T}_{\mu \nu}$ analogous to Eq.~(\ref{fcompton}) and
obtain the perturbative quantities ${\cal T}_{3}$ which can be expanded in 
powers of the renormalized coupling $\ar \equiv \as/(4\pi)$ at the scale 
$\mu^{\,2} = \Qs$,
\beq
\label{qcompton}
{\cal T}_{3}(N) \; =\; \left(
 {\cal T}^{(0)}_{3}(N) \:+\: a_{s}\, {\cal T}^{(1)}_{3}(N)
 \:+\:  a_{s}^{2}\, {\cal T}^{(2)}_{3}(N) 
 \:+\: a_{s}^{3}\, {\cal T}^{(3)}_{3}(N) \:+\: \dots \right) \: A_{\rm v}
\:\: ,
\eeq
where the $N$-independent $A_{\rm v}$ denotes the quark operator matrix element.
After performing the operator renormalization according to Eqs.~(\ref{ZA})
-- (\ref{Oren}) and (\ref{Zv3}), one thus arrives at explicit expressions for 
the~${\cal T}^{(l)}_{3}$. The anomalous dimensions $\gamma_{\,\rm v}^{\,(l)}$ 
can then be read off from the poles in $\ep$, while the coefficient function 
$C_3^{\,-}$ is related to the finite terms in $\epsilon$. During this 
procedure it is important know the expansion of the $D$-dimensional 
coefficient function 
\beq
\label{cf-exp}
 {\cal C}_{3}^{\,-}(\as,N) \:\: = \:\: \sum_{l=0}^{\infty} \: a_{\rm s}^{\, l} 
 \bigg( c_{3}^{\,(l)}(N) \:+\: \epsilon\, a_{3}^{\,(l)}(N) \:+\: 
 \epsilon^2 b_{3}^{\,(l)}(N) \:+\: \ldots \bigg)
\eeq
to a sufficient depth (positive powers) in $\ep$ at lower orders. We normalize 
the $\as^{\,0}$ contribution to Eq.~(\ref{qcompton}) including $A_{\rm v}$, 
i.e.,  
\beq
\label{T3-0}
  {\cal T}^{(0)}_{3,\pm}(N) \:\: = \:\: 1
\eeq
implies 
\beq
\label{eq:c3-0}
  c^{(0)}_{3,\pm}(N) \:\: = \:\: 1  \:\: , \quad
  a^{(0)}_{3,\pm}(N) \:\: = \:\: b^{(0)}_{3,\pm}(N) 
  \:\: = \:\: \ldots \:\: = \:\: 0   
\:\: .
\eeq

In this normalization also the first-order expression is the same for 
$F_{3}^{\,W^++W^-\!}$ and $F_{3}^{\,W^+-W^-\!}$. Again suppressing the 
$N$-dependence it reads
\bea
  \label{eq:T3-1}
  {\cal T}^{(1)}_{3,\pm} &\! =\!& 
   \frac{1}{\ep}\: \gamma^{\,(0)}_{\,\rm qq} 
   \:+\: c^{(1)}_{3,\pm} \:+\: \ep\, a^{(1)}_{3,\pm} 
   \:+\: \ep^2 b^{(1)}_{3,\pm} + {\cal O}(\ep^{3})
   \:\: . 
\eea
The corresponding second- and third-order results are given by 
\bea
\label{T3-2}
\label{eq:T2n2}
  {\cal T}^{(2)}_{3,-} & \!=\! & \frac{1}{2\ep^2}\,
  \bigg\{ \left( \gqqz - \beta_0 \right) \gqqz \bigg\}
  \: + \: \frac{1}{2\ep}\, \left\{ \gnso + 2\, \ctqo\, \gqqz \right\}
\nn \\[1mm] & & \mbox{}
  \: + \: \ctnt + \atqo\, \gqqz
  \: + \: \ep\, \bigg\{ \atnt + \btqo\, \gqqz \bigg\} 
\eea
and 
\bea
\label{T3-3}
  {\cal T}^{(3)}_{3,-} & \!=\! & \frac{1}{6\ep^3}\,
  \bigg\{ \left( \gqqz - 2\beta_0 \right)
    \left( \gqqz - \beta_0 \right) \gqqz \bigg\}
\nn \\[1mm] & & \hspn\hspn\:\mbox{}
  \: + \: \frac{1}{6\ep^2}\, \bigg\{ 3 \gnso\,\gqqz - 2 \beta_0\,\gnso
     - 2\beta_1\,\gqqz + 3\ctqo \left( \gqqz - \beta_0 \right) \gqqz
       \bigg\}
\nn \\[1mm] & & \hspn\hspn\:\mbox{}
  \: + \: \frac{1}{6\ep}\;\bigg\{ 2 \gnst + 3 \ctqo\,\gnso + 6 \ctnt\,
     \gqqz + 3 \atqo \left( \gqqz - \beta_0 \right) \gqqz \bigg\}
\nn \\[1mm] & & \hspn\hspn\:\mbox{}
  \: + \: \ctnd + \frac{1}{2}\, \atqo\,\gnso + \atnt\,\gqqz +
     \frac{1}{2}\, \btqo \left( \gqqz - \beta_0 \right) \gqqz \:\: .
\eea
Consequently the two- and three-loop coefficient functions can be read off
from the $\ep$-indepen\-dent parts of ${\cal T}^{(2)}_{3,-}$ and 
${\cal T}^{(3)}_{3,-}$ after subtracting the respective contributions due to 
the lower-order $\ep$ and $\ep^2$ quantities in Eq.~(\ref{cf-exp}). 

We are now in a position to give the comment announced above on the $\ep$-terms
in Eq.~(\ref{Z5}). These terms do not affect the extracted coefficient 
functions $\ctnt$ and $\ctnd$. Without them, however, the functions $\atqo$, 
$\btqo$ and $\atnt$ would exhibit an unphysical behaviour of their 
$N$-independent terms (which should be the same as those for $\Ftwo$), a 
feature irrelevant here but unwanted for more general applications such as the 
determination of time-like (fragmentation) coefficient functions via a suitable 
analytic continuation of the DIS results, cf.~Refs.~\cite
{Mitov:2006ic,Moch:2007tx}. In other words, Eq.~(\ref{Z5}) effectively restores
the anticommutativity of $\gamma_{\,5}$ also for those $\ep$ and $\ep^2$ 
contributions. \pagebreak

The actual computation of the Feynman diagrams for the contributions to
Eq.~(\ref{qcompton}) follows those of Refs.~\cite
{Moch:2004pa,Vogt:2004mw,Moch:2004xu,Vermaseren:2005qc,Moch:2002sn} in every
respect, so we can be very brief here. The graphs have been generated 
automatically with the diagram generator {\sc Qgraf}~\cite{Nogueira:1991ex}.
All further symbolic manipulations have been performed in {\sc Form}~\cite
{Vermaseren:2000nd,Vermaseren:2002rp}, using the {\sc Summer} package~\cite
{Vermaseren:1998uu} for the analytic evaluation of all nested sums. The
calculation relied on a massive tabulation of intermediate integrals, and 
check of all intermediate and final results were performed for fixed values of
$N$ against the {\sc Mincer} program~\cite{Gorishnii:1989gt,Larin:1991fz} and
the findings to Ref.~\cite{Retey:2000nq}. The three-loop $1/\ep\,$ terms of the
present calculation have already been used to complete the set of non-singlet 
NNLO splitting functions~\cite{Moch:2004pa}.
 
There are two aspects which deserve special attention in the context of the 
present calculation.
The first is the appearance of two functions $\gfunct1$, $\gfunct2$ in the 
final odd-$N$ results for the coefficient functions (see Refs.~\cite
{Vermaseren:2005qc} for very similar even-$N$ functions in the 
results of $\Ftwo$ and $\FL$) which fall outside the class of simple harmonic 
sums \cite{Vermaseren:1998uu} sufficient at previous orders,
\beq
\label{g12N}
g_n(N) \; = \; 
        N^{\:\!n} \: 
        \Big(
         5\:\! \* \z5
       - 2\, \* \S(-5)
       + 4\, \* \S(-2) \* \z3
       - 4\, \* \Ss(-2,-3)
       + 8\, \* \Sss(-2,-2,1)
       + 4\, \* \Ss(3,-2)
       - 4\, \* \Ss(4,1)
       + 2\, \* \S(5)
       \Big)
\:\: .
\eeq
Note that the bracketed combination of harmonic sums vanishes as $1/N^{\,2}$ 
for $N \to \infty$, hence $g_n(N)$ do not contribute to the leading 
$\,\ln^{\,k} N$, $\,k = 1,\,\ldots ,\, 6$, behaviour of the coefficient 
function $c^{(3)}_{3,-}(N)$. The $x$-space expressions corresponding to 
Eqs.~(\ref{g12N}) can be found at the end of Appendix A.
  
\begin{figure}[htb]
\vspace{2mm}
\centerline{\epsfig{file=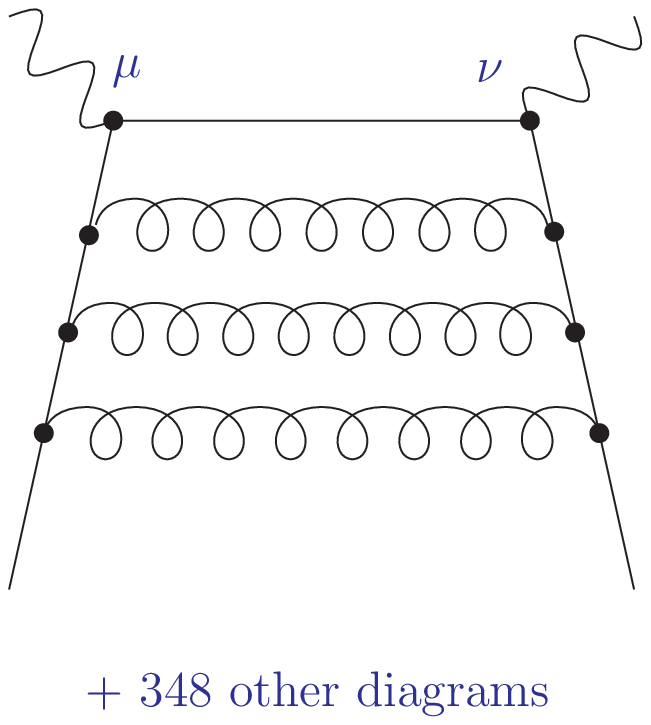,width=6.2cm}\hspace*{12mm}
\epsfig{file=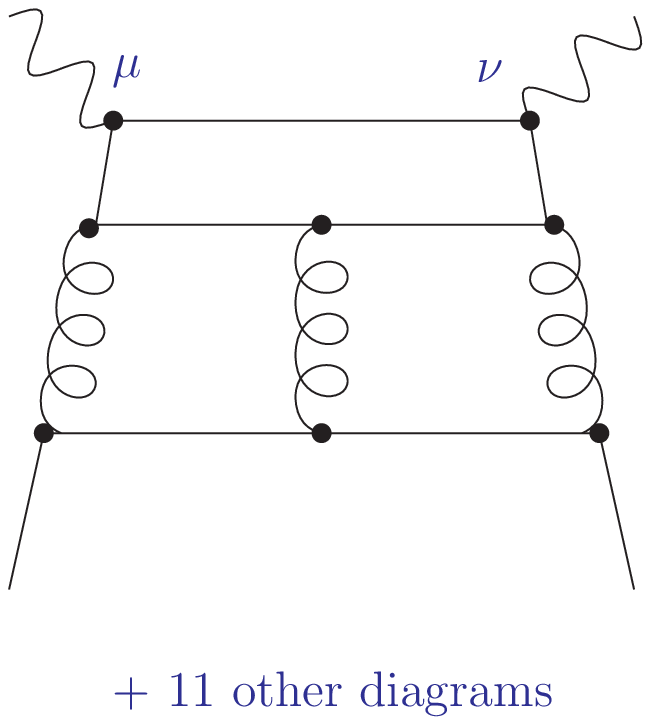,width=6.2cm}}
\vspace*{-2mm}
\caption[]{\label{fig:dabc-diags}
Typical Feynman diagrams for the contributions without (left) and with (right)
$\dabc2$ to the charged-current DIS coefficient function
$C_{3}^{\,-}$ up to the third order in the strong coupling $\alpha_{\rm s}$.}
\vspace*{-3mm}
\end{figure}

The second interesting point is the presence of a new colour structure, a 
contribution proportional to the higher SU($n_c$) group invariant $\dabc2$ 
in the three-loop coefficient function $c_{3,-}^{(3)}$.
This new colour factor is related to Feynman diagrams of a particular topology
concerning the fermion flow through the diagram. We distinguish the two cases 
(so-called flavour classes) $fl_2$ and $\fl02$ depending on whether both 
$W$-bosons are attached to the open fermion line of the initial and final 
state quark ($fl_2$) or whether both gauge bosons are coupled to a closed 
fermion loop ($\fl02$).
Sample diagrams for the two cases are displayed in Fig.~\ref{fig:dabc-diags}.
The impact of the new $\fl02$ contributions with $\,\dabc2$ on the 
coefficient function will be discussed in the next sections.
%
%
\setcounter{equation}{0}
\section{Results and discussion}
\label{sec:results}
%
%
We are now ready to present the charged-current coefficient function 
$C_3^{\,-}$ to the third order in the strong coupling $\,\ar = \as/(4\pi)$, 
i.e., the coefficients $c_{3,-}^{\,(l\leq 3)}$ in
\beq
\label{C3min} 
  F_3^{\,W^+ + W^-} \; = \; C_3^{\,-} \,\otimes\, q_{\rm val} \; = \; 
  \left( \delta(1-x) \,+\, \ar\, c_{3,-}^{\,(1)} 
  \,+\, \ar^{\,2}\, c_{3,-}^{\,(2)} \,+\, \ar^{\,3}\, c_{3,-}^{\,(3)} 
  \,+\, \ldots \right) \,\otimes\, q_{\rm val} \; .
\eeq
Here $q_{\rm val}$ represents the total (flavour summed) valence quark
distribution of the hadron, $q_{\rm val} = \sum_{\,i=1}^{\,\nf} (q_i^{} - 
\bar{q}_i^{})$ where $\nf$ is the number of effectively massless flavours.
$\otimes$ denotes the Mellin convolution which turns into a simple 
multiplication in $N$-space. All results below will be given in the \MSb\ 
scheme for the standard choice $\mu_{\:\!\rm r}^{\,2} = \mu_{\:\!\rm f}^{\,2} = 
\Qs$ of the renormalization and factorization scales.
The complete expressions for the dependence on $\mu_{\:\!\rm r}^{}$ and 
$\mu_{\:\!\rm f}^{}$ to the third order can be found, for example, in 
Eqs.~(2.16) -- (2.18) of Ref.~\cite{vanNeerven:2000uj}.

As discussed above, our calculation via the optical theorem and a dispersion 
relation directly determines the coefficient function (\ref{C3min}) for all 
odd-integer moments $N$ in terms of harmonic sums 
\cite{Gonzalez-Arroyo:1979df,Vermaseren:1998uu,Blumlein:1998if}. From these 
functions the $x$-space expressions can be reconstructed algebraically 
\mbox{\cite{Moch:1999eb,Remiddi:1999ew}} in terms of harmonic polylogarithms~%
\cite{Remiddi:1999ew,Goncharov,Borwein}. 
As in the case of $F_{\:\! 2,L}$ in electromagnetic DIS presented before 
\cite{Moch:2004xu,Vermaseren:2005qc}, the exact third-order expressions are 
unpleasantly long in both \mbox{$N$-space} and \mbox{$x$-space}. We therefore 
refrain from writing down the former in this article, and defer the latter to 
Appendix~A.

For the convenience of the reader we first recall the known results up to the 
second order. The first-order contribution to Eq.~(\ref{C3min}) is identical
to that for the $W^+\!-\! W^-$ case $C_3^{\,+}$ and given by
\cite{Bardeen:1978yd} 
\bea
\label{c3ns1}
  c_{3,\pm}^{\,(1)}(x) &\! =\! &
  \cf \{ 4\, \DD1 - 3\, \DD0 - (9 + 4\,\z2)\,\delta(x_1^{})
  - 2\,(1+x) (L_1 - L_0)
  \nn \\ & & \mbox{} \quad
  - 4\, x_1^{-1} L_0 + 4 + 2\,x \}
\eea
with $\cf = (n_c^{\,2}-1)/(2n_c) = 4/3$ and $\,n_c = \ca = 3$ in QCD. 
Here and below we use the abbreviations
\beq
\label{abbrev}
  x_1^{} \: = \: 1-x            \:\: ,\quad
  L_0 \: = \: \ln\, x        \:\: ,\quad
  L_1 \: = \: \ln\, x_1^{}      \:\: ,\quad
  \DDk \: = \: [\, x_1^{-1} L_1^{\:\!k\,}]_+ \:\: .
\eeq
As usual, the +-distributions are defined via
\beq
\label{plus}
  \int_0^1 \! dx \: a(x)_+ f(x) \; = \; \int_0^1 \! dx \: a(x)
  \,\{ f(x) - f(1) \}
\eeq
for regular functions $f(x)$. Convolutions with the distributions $\DDk$ in 
Eq.~(\ref{abbrev}) can be written as
\beq
\label{Dkconv}
  x[\DDk \otimes f](x) \; = \; 
  \int_x^1 \! dy \; \frac{\ln^{\,k} (1-y)}{1-y} \left\{ \frac{x}{y}\,
  f\!\left( \frac{x}{y} \right) - xf(x) \right\}
  \, + \: xf(x) \; \frac{\ln^{\,k+1}(1-x)}{k+1} \:\: .
\eeq

Already at two loops the coefficient functions involve polylogarithms and
Nielsen functions, hence it is convenient to have at one's disposal accurate 
parametrizations in terms of the elementary functions in Eq.~(\ref{abbrev}).
After inserting the QCD values of the colour factors $\cf$ and $\ca$, the exact
two-loop coefficient function (A.3) first obtained in 
Refs.~\cite{Zijlstra:1992kj,Moch:1999eb} can be represented by \pagebreak
\bea
\label{c3ns2}
  c_{\,3,-}^{\,(2)}(x)\!\! & \cong\!\! &  \quad
       128/9\: \DD3 - 184/3\: \DD2 - 31.1052\: \DD1 + 188.641\: \DD0
     - 338.572\: \delta (x_1^{})
  \nn \\ & & \mbox{} \quad
     - 16.40\: L_1^3 + 78.46\: L_1^2 - 470.6\: L_1
     - 149.75 - 693.2\: x + 0.218\: x L_0^4
  \nn \\ & & \mbox{} \quad
     + L_0 L_1 ( 33.62\: L_0 - 117.8\: L_1)
     - 49.30\: L_0 - 94/3\: L_0^2 - 104/27\: L_0^3
  \nn \\[1mm] &+\!\!& \mbox{} \nf \:\big\{ \,
       16/9\: \DD2 - 232/27\: \DD1 + 6.34888\: \DD0
     + 46.8464\: \delta (x_1^{}) + 0.066\: L_1^3 
  \nn \\ & & \mbox{} \quad
     - 0.663\: L_1^2 + 24.86\: L_1 - 5.738 - 5.845\: x 
     - 10.235\: x^2 - 0.190\: xL_0^3
  \nn \\ & & \mbox{} \quad
     + 4.265\: L_0 L_1 + 20/9\: L_0 ( 4 + L_0 ) \,\big\}
\eea
with an error well below 0.1\% at all values of $x$. This expression is 
slightly less compact, but considerably more accurate than the (practically 
sufficient) previous parametrization in Ref.~\cite{vanNeerven:1999ca}. 
At this order the coefficient functions for the $W^+\!+\! W^-$ and 
$W^+\!-\! W^-$cases are different. The two-loop coefficient function 
$c_{\,3,+}^{\,(2)}$ for the latter can be evaluated via Eq.~(\ref{c3ns2}) above
and Eq.~(2.9) in Ref.~\cite{Moch:2007rq} for $\,\delta\:\! c_{\,3}^{\,(2)} 
\equiv\, c_{\,3,+}^{\,(2)} - c_{\,3,-\,}^{\,(2)}$. 

The coefficients of the +-distributions $\DDk$ in Eq.~(\ref{c3ns2}) and 
Eq.~(\ref{c3ns3}) below are exact up to a truncation of the values of the 
Riemann $\zeta$-function. Also exact 
are those coefficients of $L_0^k \equiv\ln^{\,k} x$ and $L_1^k \equiv \ln^{\,k}
(1-x)$ given as fractions. Most of the remaining coefficients have been 
obtained by fits to the exact coefficient functions at $10^{-6} \leq x \leq 1\!
-\! 10^{-6}$. 
Finally the coefficients of $\delta (1-x)$ have been adjusted very slightly 
from their exact values using the lowest integer moments, thus fine-tuning the
convolution with the quark distributions to maximal accuracy (cf.~the 
discussion at the end of Section 4 of Ref.~\cite{Vogt:2004mw}. 

We now turn to our new three-loop results. Inserting, as in Eq.~(\ref{c3ns2})
above, the numerical QCD values of the $\nf$-independent SU($n_c$) colour 
factors, the third-order contribution (A.4) to Eq.~(\ref{C3min}) can be 
approximated by
\bea
\label{c3ns3}
  c_{\,3,-}^{\,(3)}(x)\!\! & \cong\!\! &  \quad
       512/27\: \DD5 - 5440/27\: \DD4 + 501.099\: \DD3
     + 1171.54\: \DD2 - 7328.45\: \DD1 
  \nn \\ && \mbox{} \quad
     + 4442.76\: \DD0 - 9172.68\: \delta (x_1^{})
     - 512/27\: L_1^5 + 8896/27\: L_1^4 - 1396\: L_1^3
  \nn \\ & & \mbox{} \quad
     + 3990\: L_1^2 + 14363\: L_1
     - 1853 - 5709\: x + x\,x_1^{} (5600 - 1432\: x)
  \nn \\ & & \mbox{} \quad
     - L_0 L_1 (4007 + 1312\: L_0) - 0.463\: xL_0^6 
     - 293.3\: L_0 - 1488\: L_0^2 - 496.95\: L_0^3 
  \nn \\ & & \mbox{} \quad
  - 4036/81\: L_0^4 - 536/405\: L_0^5
  \nn \\[1mm] &+\!\!& \mbox{} \nf \:\big\{ \,
     640/81\: \DD4 - 6592/81\: \DD3 + 220.573\: \DD2
     + 294.906\: \DD1 - 729.359\: \DD0 
  \nn \\ & & \mbox{} \quad
     + 2575.46\: \delta (x_1^{}) - 640/81\: L_1^4 + 32576/243\: L_1^3 
     - 660.7\: L_1^2 + 959.1\: L_1 
  \nn \\ & & \mbox{} \quad
     + 516.1 - 465.2\: x + x\,x_1^{} (635.3 + 310.4\: x)
     + 31.95\: x_1^{} L_1^4
  \nn \\ & & \mbox{} \quad
     + L_0 L_1\, (1496 + 270.1\:L_0 - 1191\: L_1)
     - 1.200\: xL_0^4 + 366.9\: L_0
     + 305.32\: L_0^2 
  \nn \\ & & \mbox{} \quad
     + 48512/729\: L_0^3 + 304/81\: L_0^4 \, \big\}
  \nn \\[1mm] &+\!\!& \mbox{} \nf^{\!\!\! 2} \:\big\{ \,
     64/81\: \DD3 - 464/81\: \DD2 + 7.67505\: \DD1 
     + 1.00830\: \DD0
     - 103.2602\: \delta (x_1^{}) 
  \nn \\ & & \mbox{} \quad
     - 64/81\: L_1^3 + 992/81\: L_1^2 - 49.65\: L_1 
     + 11.32 
     - x\,x_1^{} (44.52 + 11.05\: x)
  \nn \\ & & \mbox{} \quad
     + 51.94\: x + 0.0647\:xL_0^4
     - L_0 L_1\, ( 39.99 + 5.103\: L_0 - 16.30\: L_1)
     - 16.00\: L_0 
  \nn \\ & & \mbox{} \quad
     - 2848/243\: L_0^2 - 368/243\: L_0^3 \, \big\}
  \nn \\[1mm] &+\!\!& \mbox{} \!\! fl_{02}\: \nf \: \big\{
     2.147\: L_1^2 - 24.57\: L_1 + 48.79 - x_1^{} (242.4 - 150.7\: x)
     - L_0 L_1\, (81.70 
  \nn \\ & & \mbox{} \quad
     + 9.412\: L_1)
     + xL_0 \, (218.1 + 82.27\,L_0^2)
     - 477.0\: L_0 - 113.4\: L_0^2  + 17.26\: L_0^3
  \nn \\ & & \mbox{} \quad
     - 16/27\: L_0^5
     \,\big\} \: x_1^{} \:\: .
\eea
Here the factor $f\:\!\!l_{02}$ ($=1$ for the numerical evaluation) indicates 
the $\dabc2$ contribution entering at this order for the first time, cf.\ 
Ref.~\cite{Larin:1991tj}.
Also Eq.~(\ref{c3ns3}), first presented in Ref.~\cite{Vogt:2006bt}, deviates by
much less than one part in a thousand from the corresponding exact expression 
which we evaluated using a weight-five extension of the {\sc Fortran} package 
\cite{Gehrmann:2001pz} for the harmonic polylogarithms. 
Eqs.~(\ref{c3ns2}) and (\ref{c3ns3}) can be readily transformed to Mellin space
at complex values of $N$ (using, e.g., the appendix of Ref.~\cite
{Blumlein:1998if} for the moments of $\,\ln x \,\ln^{\:\!2}(1-x)\:\!$ etc)
for use with $N$-space programs, such as {\sc QCD-Pegasus} \cite{Vogt:2004ns}, 
for the evolution of parton densities and structure functions.

The $x\! \ra\! 1$ and $x\! \ra\! 0$ end-point behaviour of the higher-order 
contributions to Eq.~(\ref{C3min}) is of special interest, both theoretically
and phenomenologically. We first address the large-$x$ limit. Here the leading 
terms of $c_{\,3,\pm}^{\,(n)}(x)$ are the soft-gluon +-distributions $\DDk$ 
with $k = 0,\:\ldots ,\: 2n\!-\!1$. To order $\as^3$ the corresponding 
coefficients are identical to those for the electromagnetic and charged-current
structure functions $\Ftwo$ which can be found in Eqs.~(4.14) -- (4.19) of 
Ref.~\cite{Vermaseren:2005qc}.\footnote
{$\,$The sign of the term $- 232\, \z5$ in the first line of Eq.~(4.19) has 
been misprinted in the hep-ph version of Ref.~\cite{Vermaseren:2005qc}. The 
journal version is correct.} 

The terms with $\delta (1-x)$ (arising from virtual corrections and soft-gluon
contributions) are the same for all four charged-current coefficient functions 
$c_{\,2,3,\pm}^{\,(3)}$ and differ from the corresponding photon-exchange
quantity of Ref.~\cite{Vermaseren:2005qc} only by the obvious absence of 
contributions from the $fl_{11}$ flavour classes (where the photons couple to 
different quark loops) in $W$-exchange processes. For the convenience of the
reader we here collect the third-order $\delta (1-x)$ terms scattered over the 
13~pages of Eq.~(A.4)$\:\!$: 
\bea
\label{c3delta}
  c_{\,3,\pm}^{\,(3)} \Big|_{\,\delta(x_1^{})} \!\!\! & = \! &
  \cas \* \cf \*
   \:\Bigg[
          - {1909753 \over 1944}
          - {143255 \over 81}\: \* \z2
          + {105712 \over 81}\: \* \z3
          + {25184 \over 135}\: \* \zss
          - {416 \over 3}\: \* \z5
\nn \\[0.5mm] & &
          + \: 540\, \* \z2 \* \z3
          - {248\over 3} \* \zts
          - {3512 \over 63}\: \* \zst 
  \Bigg]
  \: \: + \: \: \ca \* \cfs \* 
  \:\Bigg[
            {9161 \over 12}
          + {104117 \over 54}\: \* \z2
\nn \\[0.5mm] & & 
          - \: {6419 \over 3}\: \* \z3
          + {87632 \over 135}\: \* \zss
          - {4952 \over 9}\: \* \z5
          - {6644 \over 9}\: \* \z2 \* \z3
          + {1016 \over 3}\: \* \zts
          - {33556 \over 315}\: \* \zst
  \Bigg]
\nn \\[0.5mm] & & \mbox{\hspn}
  + \: \cft \* 
  \:\Bigg[
          - {7255 \over 24}
          - {3379 \over 6}\: \* \z2
          - 318\, \* \z3
          - {2148 \over 5}\: \* \zss
          + 1240\, \* \z5
          + 808\, \* \z2 \* \z3
\nn \\[0.5mm] & & 
          - \: {304 \over 3}\: \* \zts
          + {4184 \over 315}\: \* \zst
  \Bigg]
  \: \: + \: \: \cf\, \* \nfs \*
  \:\Bigg[
          - {9517 \over 486}
          - {860 \over 27}\: \* \z2
          - {152 \over 81}\: \* \z3
          - {32 \over 27}\: \* \zss
  \Bigg]
\nn \\[0.5mm] & & \mbox{\hspn}
  + \: \ca \* \cf\, \* \nf \* 
  \:\Bigg[
            {142883 \over 486}
          + {40862 \over 81}\: \* \z2
          - {18314 \over 81}\: \* \z3
          - {2488 \over 135}\: \* \zss
          + {8 \over 3}\: \* \z5
          - {56 \over 3}\: \* \z2 \* \z3
  \Bigg]
\nn \\[0.5mm] & & \mbox{\hspn}
  + \: \cfs\, \* \nf \*
  \:\Bigg[
          - {341 \over 36}
          - {5491 \over 27}\: \* \z2
          + {1348 \over 3}\: \* \z3
          - {16472 \over 135}\: \* \zss
          - {592 \over 9}\: \* \z5
          - {352 \over 9}\: \* \z2 \* \z3
  \Bigg]
\:\: . \quad
\eea
\pagebreak

Finally we consider the subleading (integrable) large-$x$ logarithms which have
attracted renewed theoretical interest recently 
\cite{Grunberg:2007nc,Laenen:2008ux,Laenen:2008gt}. 
Terms up to $\ln^{\, 2n-1} (1-x)$ occur in the $n$-th order coefficient 
functions. Their coefficients are the same for the $W^+\!+ W^-$ and 
$W^+\!-\! W^-$cases. Unlike the \mbox{+-distributions}, however, these 
contributions differ (except for the highest power in each colour factor, where
the coefficient is minus that of the highest +-distributions) between $\Ftwo$ 
and $\F3$. For the present structure function the two-loop contributions read
\bea
\label{c32L13}
  c_{\,3,\pm}^{\,(2)} \Big|_{\,L_1^3} \!\!\! & = \! &
          - \: 8\: \* \cfs
\\[0.5mm]
\label{c32L12}
  c_{\,3,\pm}^{\,(2)} \Big|_{\,L_1^2} \!\!\! & = \! &
            {22 \over 3}\: \* \ca \* \cf
       \: + \: 52\: \* \cfs
       \: - \: {4 \over 3}\: \* \cf\, \* \nf 
\\[0.5mm]
\label{c32L11}
  c_{\,3,\pm}^{\,(2)} \Big|_{\,L_1} \!\!\! & = \! &
       - \: \ca \* \cf \*  \:\Bigg[
            {640 \over 9}
          - 8\, \* \z2
         \Bigg] 
     \: - \: \cfs  \* \: [
            16
          - 32\, \* \z2
          ]
     \: + \: {124 \over 9}\: \* \cf\, \* \nf  
\:\: .
\eea
The corresponding coefficients for the third-order coefficient functions are
given by
\bea
\label{c33L15}
  c_{\,3,\pm}^{\,(3)} \Big|_{\,L_1^5} \!\!\! & = \! &
          - \: 8\: \* \cft
\\[0.5mm]
\label{c33L14}
  c_{\,3,\pm}^{\,(3)} \Big|_{\,L_1^4} \!\!\! & = \! &
            {220 \over 9} \: \* \ca \* \cfs
       \: + \: 84\: \* \cft
       \: - \: {40 \over 9} \: \* \cfs\, \* \nf
\\[1mm]
\label{c33L13}
  c_{\,3,\pm}^{\,(3)} \Big|_{\,L_1^3} \!\!\! & = \! &
       - \: {484 \over 27}\: \* \cas \* \cf
       - \: \ca \* \cfs \*  \:\Bigg[
            {9056 \over 27}
          - 32\, \* \z2
         \Bigg]
     \: - \: \cft  \* \: [
            110
          - 96\, \* \z2
          ]
\nn \\[1mm] & & \mbox{\hspn}
     \: + \: {176 \over 27}\: \* \ca \* \cf\, \* \nf  
     \: + \: {1640 \over 27}\: \* \cfs\, \* \nf
     \: - \: {16 \over 27}\: \* \cf\, \* \nfs
\\[2mm]
\label{c33L12}
  c_{\,3,\pm}^{\,(3)} \Big|_{\,L_1^2} \!\!\! & = \! &
          \cas \* \cf \* \:\Bigg[
            {7580 \over 27}
          - {98 \over 3}\: \* \z2
         \Bigg]
     \: + \: \ca \* \cfs \* \:\Bigg[ 
            {12031 \over 9}
          - 372\, \* \z2
          - 240\, \* \z3
          \Bigg]
\nn\\[1mm] & & \mbox{\hspn}
     - \: \cft \* \:\Bigg[
            {1097 \over 3}
          + 656\, \* \z2
          + 16\, \* \z3
          \Bigg]
     \: - \: \ca \* \cf\, \* \nf \* \:\Bigg[ 
            {2734 \over 27}
          - {16 \over 3}\: \* \z2
          \Bigg]
\nn\\[1mm] & & \mbox{\hspn}
     - \: \cfs\, \* \nf \* \:\Bigg[
            {2098 \over 9}
          - {112 \over 3}\, \* \z2
          \Bigg]
     \: + \: {248 \over 27}\: \* \cf\, \* \nfs
\\[2mm]
\label{c33L11}
  c_{\,3,\pm}^{\,(3)} \Big|_{\,L_1} \!\!\! & = \! &
        - \: \cas \* \cf \* \:\Bigg[
            {138598 \over 81}
          - {4408 \over 9}\: \* \z2
          - 272\, \* \z3
          + {176 \over 5}\: \* \zss
         \Bigg]
     \: - \: \ca \* \cfs \* \:\Bigg[
            {69833 \over 162}
\nn\\[1mm] & & 
          - \: {12568 \over 9}\: \* \z2
          - {1904 \over 3}\: \* \z3
          + {764 \over 5}\: \* \zss
          \Bigg]
     \: + \: \cft  \*  \:\Bigg[
            {1741 \over 6}
          + {1220 \over 3}\: \* \z2
\nn\\[1mm] & & 
          + 480\, \* \z3
          - {376 \over 5}\: \* \zss
          \Bigg]
     \: + \: \ca \* \cf\, \* \nf \* \:\Bigg[
            {45260 \over 81}
          - 108\, \* \z2
          - \: 16\, \* \z3
          \Bigg]
\nn\\[1mm] & & \mbox{\hspn}
     \: + \: \cfs\, \* \nf \* \:\Bigg[
            {9763 \over 81}
          - {2224 \over 9}\: \* \z2
          - {112 \over 3}\: \* \z3
          \Bigg]
     \: - \: \cf\, \* \nfs \* \:\Bigg[
            {3520 \over 81}
          - {32 \over 9}\: \* \z2
          \Bigg]
\:\: .
\eea
The extraction of some of these coefficients from Eqs.~(A.3) and (A.4) is far 
from trivial. We therefore provide the corresponding results for $\Ftwo$, which
we did not include in Ref.~\cite{Vermaseren:2005qc}, in Appendix~B where we 
also discuss an unexpected (to us) relation between the large-$x$ coefficient 
functions.
 
We now turn to the small-$x$ limit of the second- and third-order coefficient
function in Eq.~(\ref{C3min}). The leading terms in this case are the `double-%
logarithms' $L_0^k \equiv \ln^{\,k} x$ with $ k = 1,\:\ldots ,\: 2n\!-\!1$ at 
the n-th order in $\as$. The two-loop coefficients are  
\bea
\label{c32L03}
  c_{\,3,-}^{\,(2)} \Big|_{\,L_0^3} \!\!\! & = \! &
          - \: 2\: \* \ca \* \cf \: + \: {7 \over 3}\: \* \cfs
\\[1mm]
\label{c32L02}
  c_{\,3,-}^{\,(2)} \Big|_{\,L_0^2} \!\!\! & = \! &
          - \:   {103 \over 6}\: \* \ca \* \cf
       \: + \: 21\: \* \cfs
       \: + \: {5 \over 3}\: \* \cf\, \* \nf
\\[1mm]
\label{c32L01}
  c_{\,3,-}^{\,(2)} \Big|_{\,L_0} \!\!\! & = \! &
        - \: \ca \* \cf \*  \:\Bigg[
            {122 \over 3}
          - 8\, \* \z2
         \Bigg]
     \: + \: \cfs  \* \: [
            21
          + 8\, \* \z2
          ]
     \: + \: {20 \over 3}\: \* \cf\, \* \nf
\:\: .
\eea
The results for the corresponding $\,W^+\!-\! W^-$ coefficient function 
$c_{\,3,+}^{\,(2)}$ can be obtained by combining  Eqs.~(\ref{c32L03}) -- 
(\ref{c32L01}) with Eq.~(2.7) in Ref.~\cite{Moch:2007rq}. Also that equation 
for $\,\delta\:\! c_{\,3}^{\,(2)} \equiv\, c_{\,3,+}^{\,(2)} - 
c_{\,3,-\,}^{\,(2)}$ contains terms up to $\ln^{\,3} x$, hence 
$c_{\,3,-\,}^{\,(2)}$ and $c_{\,3,+}^{\,(2)}$ already differ in the leading 
logarithm. On the other hand, the leading logarithms (but only these) are the 
same for the even-$N$ based two-loop coefficient functions $c_{\,2,+}^{\,(2)}$ 
and $c_{\,3,+}^{\,(2)}$ for $\Ftwo$ and $\F3$.

At the third order in $\as$, the coefficients of $\ln^{\,k} x$ in 
Eq.~(\ref{C3min}) are given by
\bea
\label{c33L05}
  c_{\,3,-}^{\,(3)} \Big|_{\,L_0^5} \!\!\! & = \! &
          + \: {2 \over 5}\: \* \cas \* \cf
       \: - \: {29 \over 15}\: \* \ca \* \cfs
       \: + \: {53 \over 30}\: \* \cft
       \: - \: {32 \over 15}\: \* \dabcnc
\\[1mm]
\label{c33L04}
  c_{\,3,-}^{\,(3)} \Big|_{\,L_0^4} \!\!\! & = \! &
          - \: {166 \over 27}\: \* \cas \* \cf
       \: + \: {43 \over 36}\: \* \ca \* \cfs
       \: + \: {89 \over 12}\: \* \cft
       \: + \: {46 \over 27}\: \* \ca \* \cf\, \* \nf
       \: - \: {31 \over 18}\: \* \cfs \, \* \nf
\\[1mm]
\label{c33L03}
  c_{\,3,-}^{\,(3)} \Big|_{\,L_0^3} \!\!\! & = \! &
        - \: \cas \* \cf \* \:\Bigg[
            {9799 \over 81}
          + {64 \over 9}\: \* \z2
         \Bigg]
     \: + \: \ca \* \cfs \* \:\Bigg[
            {24245 \over 162}
          + {220 \over 3}\: \* \z2
          \Bigg]
\nn\\[1mm] & & \mbox{\hspn}
     - \: \cft \* \:\Bigg[
            {49 \over 3}
          + {710 \over 9}\: \* \z2
          \Bigg]
     \: + \: {772 \over 27}\: \* \ca \* \cf\, \* \nf 
     \: - \: {2179 \over 81}\: \* \cfs\, \* \nf
     \: - \: {92 \over 81}\: \* \cf\, \* \nfs
\nn\\[1mm] & & \mbox{\hspn}
     - \: \dabcnc \* \:\Bigg[
            {704 \over 9}
          - {256 \over 3}\, \* \z2
          \Bigg]
\\[2mm]
\label{c33L02}
  c_{\,3,-}^{\,(3)} \Big|_{\,L_0^2} \!\!\! & = \! &
        - \: \cas \* \cf \* \:\Bigg[
            {38642 \over 81}
          - {226 \over 9}\: \* \z2
          + {212 \over 3}\: \* \z3
         \Bigg]
     \: + \: \ca \* \cf\, \* \nf \* \:\Bigg[ 
            {4216 \over 27}
          - {112 \over 9}\: \* \z2
          \Bigg]
\nn\\[1mm] & & \mbox{\hspn}
     \: + \: \ca \* \cfs \* \:\Bigg[ 
            {13297 \over 54}
          + 401\, \* \z2
          + 328\, \* \z3
          \Bigg]
     - \: \cfs\, \* \nf \* \:\Bigg[
            {2687 \over 27}
          + {62 \over 3}\: \* \z2
          \Bigg]
\nn\\[1mm] & & \mbox{\hspn}
     + \: \cft \* \:\Bigg[
            {1085 \over 6}
          - {1373 \over 3}\: \* \z2
          - 286\, \* \z3
          \Bigg]
     \: - \: {712 \over 81}\: \* \cf\, \* \nfs
\nn\\[1mm] & & \mbox{\hspn}
     \: - \: \dabcnc \* \:\Bigg[
            1056
          - 96\, \* \z2
          - {1216 \over 3}\: \* \z3
          \Bigg]
\\[2mm]
\label{c33L01}
  c_{\,3,-}^{\,(3)} \Big|_{\,L_0} \!\!\! & = \! &
        - \: \cas \* \cf \* \:\Bigg[
            {53650 \over 81}\:
          - {11260 \over 27}\: \* \z2
          + {1696 \over 9}\: \* \z3
          + {416 \over 5}\: \* \zss
         \Bigg]
     \: - \: \ca \* \cfs \* \:\Bigg[
            {404119 \over 324}
\nn\\[0.5mm] & & 
          - {16273 \over 27}\: \* \z2
          - {9206 \over 9}\: \* \z3
          - {3416 \over 15}\: \* \zss
          \Bigg]
     \: + \: \cft  \*  \:\Bigg[
            {5553 \over 4}
          - {2308 \over 3}\: \* \z2
\nn\\[0.5mm] & & 
          - {1978 \over 3}\: \* \z3
          - {4748 \over 15}\: \* \zss
          \Bigg]
     \: + \: \ca \* \cf\, \* \nf \* \:\Bigg[
            {26084 \over 81}
          - {3112 \over 27}\: \* \z2
          + {304 \over 9}\: \* \z3
          \Bigg]
\nn\\[0.5mm] & & \mbox{\hspn}
     \: - \: \cfs\, \* \nf \* \:\Bigg[
            {271 \over 162}
          + {718 \over 27}\: \* \z2
          + {1028 \over 9}\: \* \z3
          \Bigg]
     \: - \: \cf\, \* \nfs \* \:\Bigg[
            {1684 \over 81}
          - {16 \over 3}\: \* \z2
          \Bigg]
\nn\\[0.5mm] & & \mbox{\hspn}
     \: - \: \dabcnc \* \:\Bigg[
            1600
          - {1024 \over 3}\: \* \z2
          - {1088 \over 3}\: \* \z3
          + {2176 \over 5}\: \* \zss
          \Bigg]
\:\: .
\eea
Inserting $C_A=3$, $C_F=4/3$, $\dabc2/n_c = 5\,\nf /18$ and the numerical 
values of the $\zeta$-function, Eqs.~(\ref{c33L05}) -- (\ref{c33L01}) yield
\bea
\label{c33L0num}
  c_{\,3,-}^{\,(3)} \big|_{\,L_0^5} \! & \cong \! &
    - 1.32346 - 0.59259\:\nf\: \fl02 \nn\\
  c_{\,3,-}^{\,(3)} \big|_{\,L_0^4} \! & \cong \! &
    - 49.8272 + 3.75309\:\nf \nn\\
  c_{\,3,-}^{\,(3)} \big|_{\,L_0^3} \! & \cong \! &
    - 496.842 + 66.5460\:\nf  - 1.51440\:\nfs 
    + 17.2626\:\nf\:\fl02 \nn\\
  c_{\,3,-}^{\,(3)} \big|_{\,L_0^2} \! & \cong \! &
    - 1485.12 + 305.353\:\nf - 11.7202\:\nfs 
    - 114.125\:\nf\:\fl02 \nn\\
  c_{\,3,-}^{\,(3)} \big|_{\,L_0^1} \! & \cong \! &
    - 274.294 + 367.303\:\nf - 160.229 \:\nfs 
    - 494.486\:\nf\:\fl02 
\eea
where, as in Eq.~(\ref{c3ns2}) above, $\fl02$ marks the $\dabc2$ contribution.

The qualitative pattern in Eq.~(\ref{c33L0num}) is the same as found for the 
three-loop splitting functions in Eqs.~(4.16) and (4.18) of 
Ref.~\cite{Moch:2004pa} and the non-singlet photon-exchange coefficient 
function for $\Ftwo$ in Eq.~(4.26) of Ref.~\cite{Vermaseren:2005qc}$\:\!$: 
The coefficients show a dramatic rise
until the fourth term, thus precluding any meaningful approximation by (up to)
the three highest contributions (\ref{c33L05}) -- (\ref{c33L03}) at practically
relevant values of $x$. It is also worthwhile to note that the new $\dabc2$ 
colour structure actually forms the largest contribution to the first line of
Eq.~(\ref{c3ns2}) for physical values of the number of flavours. Hence it is 
not possible to obtain, or reliably estimate, even the leading $\as^{\,n} 
\ln^{\,2n-1}x$ terms at higher orders $n$ by any resummation of lower-order 
structures.
 
For the corresponding $\,W^+\!-\! W^-$ coefficient function $c_{\,3,+}^{\,(3)}$ 
only the six lowest even-integer moments have been computed so far \cite
{Moch:2007gx,Rogal:2007bv}. As observed below Eq.~(\ref{c32L01}), the 
leading-log coefficients are the same for the even-$N$ coefficient functions 
for  $\Ftwo$ and $\F3$ to two loops. Assuming that this relations holds at 
three loops as well, we can employ Eq.~(\ref{c33L05}) above and Eq.~(4.14) in 
Ref.~\cite{Vermaseren:2005qc} to derive the conjecture
\beq
\label{c33dfL05}
 \delta\:\! c_{\,3}^{\,(3)} \Big|_{\,L_0^5} \; = \;
         {1 \over 3}\: (\ca - 2\,\cf)\, \cfs 
 \: - \: {2 \over 5}\: (\ca - 2\,\cf)^2\, \cf \:\: .
\eeq 
The relation is supported by the fact that, as the even-integer moments 
\cite{Moch:2007rq} of \mbox{$\,\delta\:\! c_{\,3}^{\,(3)} \equiv\, 
c_{\,3,+}^{\,(3)} - c_{\,3,-\,}^{\,(3)}$} (with the $C_3^{\,-}$-specific 
$\dabc2$ contribution removed from the latter quantity), also the coefficient 
(\ref{c33dfL05}) shows the characteristic $1/n_c^{\,2}$ suppression in the 
limit of a large number of colours $n_c$ predicted in 
Ref.~\cite{Broadhurst:2004jx}. We expect to be able to verify the above 
conjecture by a complete calculation of the coefficient function 
$c_{\,3,+}^{\,(3)}(x)\,$ in the not too distant future.
%
%
\setcounter{equation}{0}
\section{Numerical implications}
\label{sec:numerics}
%
%
We start this section by graphically illustrating our new third-order
coefficient function $c_{\,3,-}^{\,(3)}(x)$ and its convolution with a
schematic but sufficiently characteristic quark distribution,
\beq
\label{shape}
   xf(x) \:\: = \:\: x^{\,0.5} (1-x)^3 \; .
\eeq
All curves in the corresponding two figures are scaled down from the 
normalization (\ref{C3min}) by a factor $2000 \simeq (4\pi)^3$, effectively 
switching back to the normal-sized expansion parameter $\as\,$.

\begin{figure}[bht]
\centerline{\epsfig{file=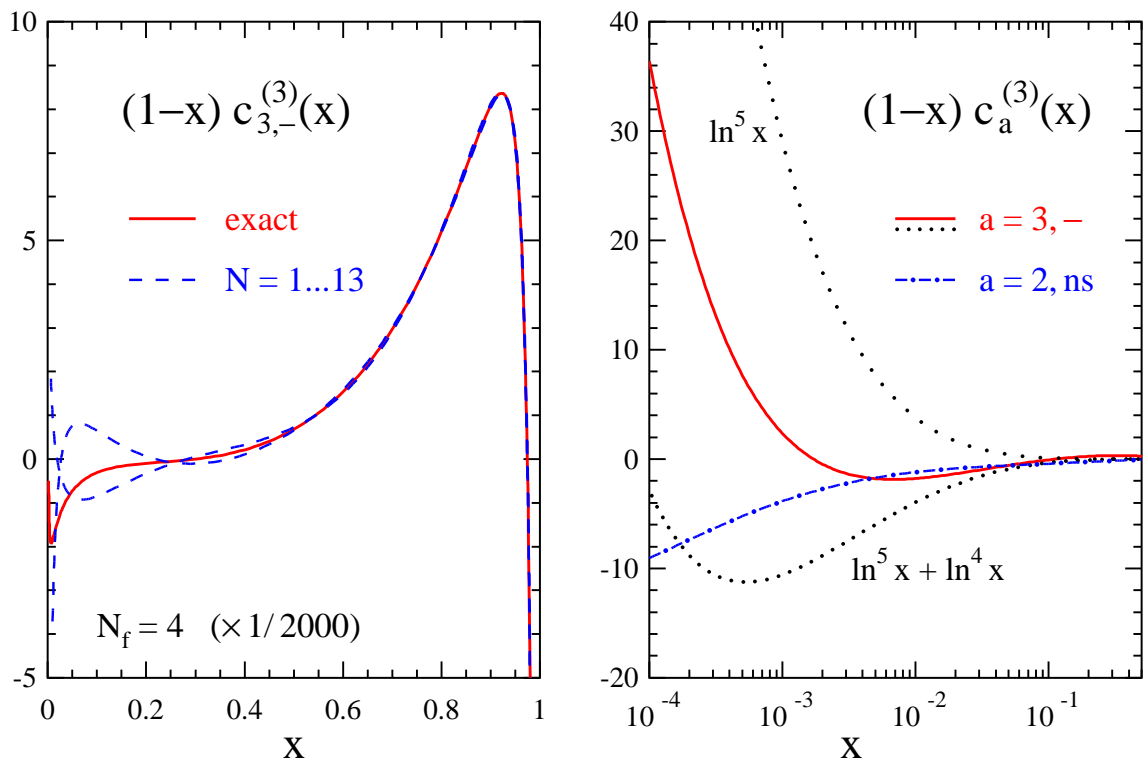,width=16.0cm}\quad}
\vspace{-1mm}
\caption{ \label{pic:c3m3}
 The third-order coefficient function $c_{\,3,-}^{\,(3)}(x)$ for four flavours,
 multiplied by $(1\!-\!x)/2000$ for display purposes.
 Also shown (left) are the previous uncertainty band~\cite{vanNeerven:2001pe}
 and (right) the corresponding contribution to the $\,W^+\!+\! W^-\!$ structure 
 function $F_{\:\!2,\rm ns}$ (from Ref.~\cite{Vermaseren:2005qc} for 
 $fl_{11} = 0$) and the small-$x$ approximations by the leading and 
 next-to-leading logarithms (\ref{c33L05}) and (\ref{c33L04}).}
\end{figure}

On the left side of Fig.~\ref{pic:c3m3} we compare the exact $x$-shape given by
Eqs.~(\ref{c3ns3}) and (A.4) to two approximations indicating the previous
uncertainty band \cite{vanNeerven:2001pe}. This band was constrained by the
lowest seven odd-integer moments $N = 1\ldots 13$ computed in Ref.~\cite
{Retey:2000nq} and the four leading \mbox{+-distributions}~\cite{Vogt:1999xa}
fixed by the next-to-leading logarithmic threshold resummation matched
to the second-order coefficient function of Ref.~\cite{Zijlstra:1992kj}.
We see that this estimate was indeed reliable, and sufficiently accurate at
$x \gsim 0.2$. Thus a future determination of the fourth-order effects can be
started reliably from fixed-$N$ moments combined with the seven leading \mbox
{+-distributions} now fixed by the next-to-next-to-next-to-leading logarithmic
threshold resummation performed in Ref.~\cite{Moch:2005ba}. We refer the reader
to this article for a discussion of higher-order effects at very large values
of $x$.

The right side of Fig.~\ref{pic:c3m3} focuses on medium to small values of $x$.
A huge low-$x$ rise is found from $x\,\simeq10^{-3}$. This is totally different 
from the behaviour of the corresponding electromagnetic and $\,W^+\!+ W^-\!$ 
coefficient function for $\Ftwo$ also shown in the figure. In fact, this rise 
can be attributed entirely to the new $\dabc2$ colour structure in 
Eqs.~(\ref{c3ns3}) and (\ref{c33L0num}): If this contribution were removed from
the coefficient function, then the solid line (red in the archive version) 
would not end at 36.4 for $x = 10^{-4}$, but at $-4.3$, in qualitative 
agreement with the behaviour of $c_{\,2,\rm ns}^{\,(3)}(x)$.

\begin{figure}[hbt]
\vspace{-0.5mm}
\centerline{\epsfig{file=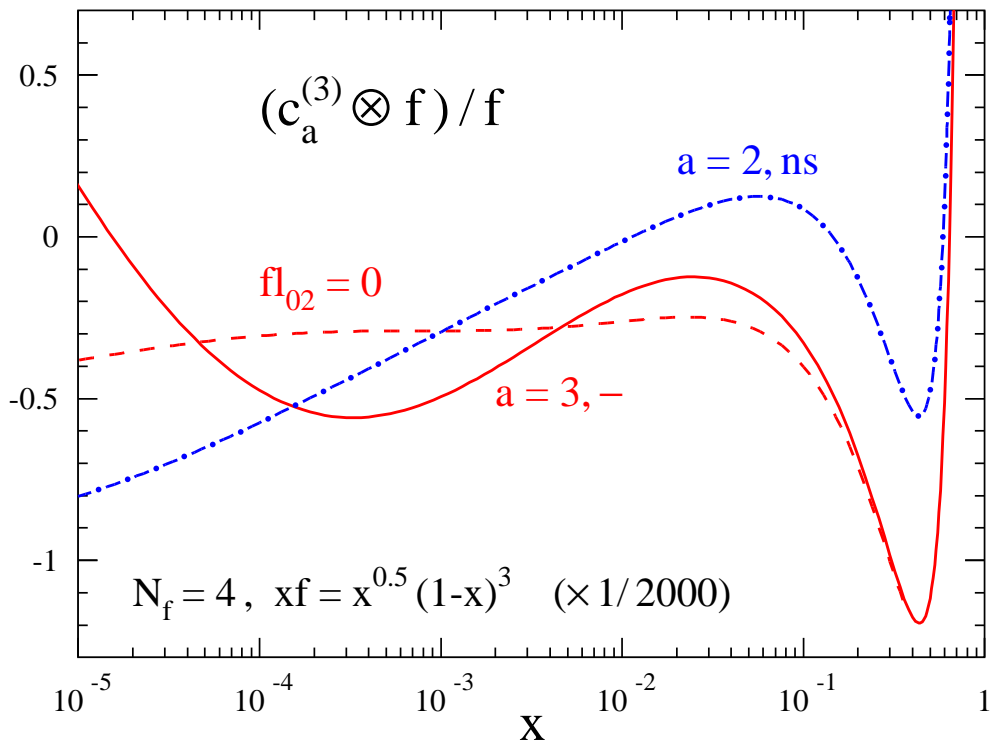,width=13.6cm}\quad}
\vspace{-1.5mm}
\caption{ \label{pic:c23cnv}
 Convolution of the $\,W^+\!+\! W^-$ charged-current coefficient functions
 $c_{\,3,-}^{\,(3)}$ and $c_{\,2,\rm ns}^{\,(3)}$ for $\nf = 4$ with a simple
 shape typical of non-singlet quark distributions of the nucleon. Also shown
 (dashed) is the result obtained for the former if the $\dabc2$ part of
 Eqs.~(\ref{c3ns3}) and (A.4) is left out. }
\end{figure}

What physically matters, of course, is not the distribution (aka generalized
function) $c_{\,3,-}^{\,(3)}(x)$ itself but the resulting contribution to the 
structure function (\ref{C3min}), obtained by the convolution (\ref{Dkconv}) 
and its obvious counterpart for the regular terms with the valence quark 
distribution. This convolution is shown in Fig.~\ref{pic:c23cnv} for the 
typical quark distribution (\ref{shape}) (the normalization is irrelevant as we
display all results normalized to $f$, thus suppressing large but trivial 
variations over the chosen wide range in $x$). For the sake of a direct 
comparison the same shape is used in Fig.~\ref{pic:c23cnv} for the (different) 
quark distribution entering $F_{\:\!2,\rm ns\,}$. 

A comparison between the result for $c_{\,3,-}^{\,(3)}(x)$ and $c_{\,2,\rm ns}
^{\,(3)}(x)$ illustrates the `$x$-shifting' power of the convolution integrals:
The former coefficient function (for $\fl02 = 0$) is larger than the latter at 
all $x < 0.8$, yet its convolution result is larger only for $x < 10^{-3}$. 
Note also that the small-$x$ rise for $\F3$, already delayed in 
Fig.~\ref{pic:c3m3} by about one order of magnitude in $x$ to $x\lsim 10^{-3}$
by non-leading logarithms, is confined to $x \lsim 10^{-4}$ after the 
convolution.

The normalized convolution $(c_{\,3,-}^{\,(3)} \otimes f) / f$ of Fig.~2, i.e., 
the coefficient of $\as^3$ in $F_3^{\,W^+ + W^-}\!/q_{\rm val}$ for 
$xq_{\rm val} = x^{\,0.5} (1-x)^3$, exceeds three, thus 2.5\% of the lowest 
order for $\as \simeq 0.2$, only at $x > 0.75$ (where we run out of 
measurements far from the resonance region anyway) and $x < 1.5\cdot 10^{-8}$.
Hence the perturbative expansion for this structure function appears stable at 
all practically relevant values of $x$, including the very low values $\,x\gsim
10^{-8}$ probed at $\,\Qs \approx 10^{\,4}\: \GeV2$ by the scattering of 
ultra-high energy cosmic neutrinos, see, e.g., Ref.~\cite{Quigg:2008ab}.

We finally assemble Eqs.~(\ref{c3ns1}), (\ref{c3ns2}) and (\ref{c3ns3}) to 
illustrate the perturbative expansion (\ref{C3min}) of the structure function 
$F_3^{\,W^+ + W^-\!}$. In order to easily compare the contributions of the 
various orders, we use an order-independent value of the strong coupling, 
$\as (\mu_{\:\!\rm r}^{\,2}\!=\! \mu_{\:\!\rm f}^{\,2}\!=\! \Qs)\:=\: 0.2$,
corresponding to a scale $\,\Qs \approx 30 \ldots 50\: \GeV2$. For the same 
reason we use the same quark distribution at all orders of Eq.~(\ref{C3min}), 
starting with the model shape (\ref{shape}) already employed above.

\begin{figure}[hbt]
\vspace{-0.5mm}
\centerline{\epsfig{file=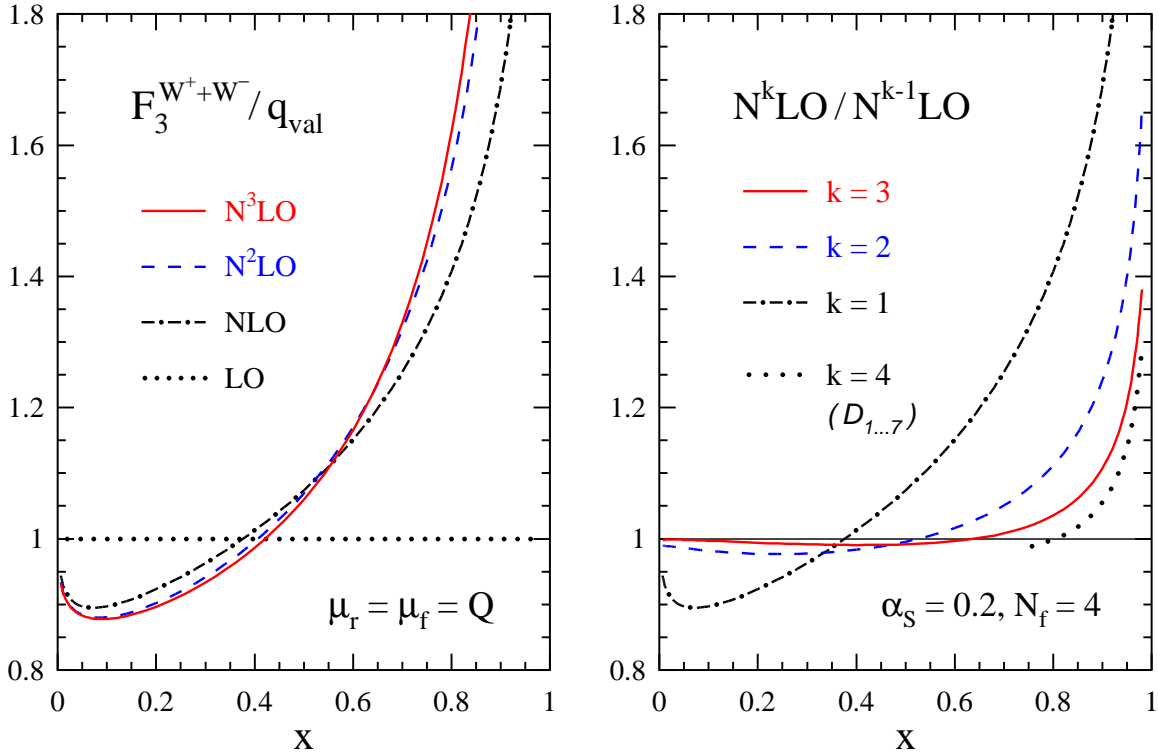,width=16.0cm}\quad}
\vspace{-1.5mm}
\caption{ \label{pic:cexpcnv1}
 The structure function (\ref{C3min}) to order $\as^{\,3}$ (N$^3$LO) for a 
 fixed value of $\as$ and the schematic valence quark distribution 
 (\ref{shape}). Next to the cumulative effect of the known corrections on the 
 left, we show on the right the relative corrections due to the $k$-loop 
 coefficient functions, including an N$^4$LO large-$x$ estimate by the 
 +-distributions ${\cal D}_{\:\!1,\ldots,7}$ known from the threshold 
 resummation \cite{Moch:2005ba}.}
\end{figure}

\begin{figure}[htp]
\centerline{\epsfig{file=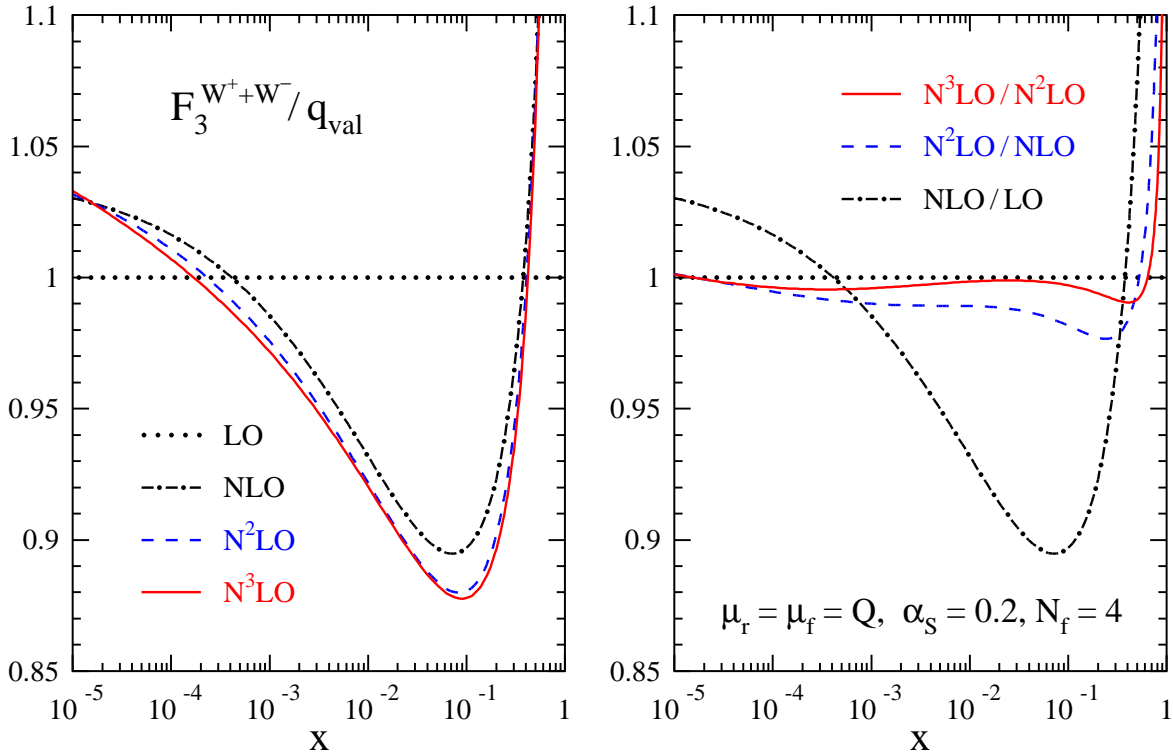,width=16cm}\quad}
\vspace{-1.5mm}
\caption{ \label{pic:cexpcnv2}
 As Fig.~\ref{pic:cexpcnv1}, but for the higher-order corrections for 
 $\,F_3^{\,W^+ + W^-}\!$ down to small values of $x$.}
\end{figure}

In Figs.~\ref{pic:cexpcnv1} and \ref{pic:cexpcnv2} we show the total results
at N$^{\,l}$LO as well as the relative effects of the \mbox{$l$-loop}
contributions to the structure function. The higher-order (NNLO and N$^3$LO)
corrections are small (1\% or less at N$^3$LO) except for the soft-gluon rise
towards $x = 1$. Note that with increasing order this rise becomes steeper, but
starts at larger values of $x$. For instance, the relative $l$-loop corrections
exceed 5\% at $x\, > 0.46,\; 0.7,\; 0.83$ for $l=1,\; 2,\; 3$, respectively.
The latter value corresponds to an invariant mass $W \simeq 3$ GeV for $\Qs =
40\: \GeV2$. 

An estimate of the large-$x$ fourth-order coefficient function by the seven
+-distributions fixed the next-to-next-to-next-to-leading logarithmic
soft-gluon (threshold) resummation \cite{Moch:2005ba} strongly suggests that
this trend will continue at even higher orders. As shown in the right part of
Fig.~\ref{pic:cexpcnv1}, this estimate leads to $x \gsim 0.9$ for a relative
four-loop correction exceeding 5\%. All in all, the N$^3$LO expansion can be
considered safe to, at least, $x \simeq 0.8$ at $\Qs\approx 40\: \GeV2$
(the safe range, of course, widens (shrinks) with increasing (decreasing)
$\Qs$ due to the scale dependence of $\as$).

\begin{figure}[htp]
\centerline{\epsfig{file=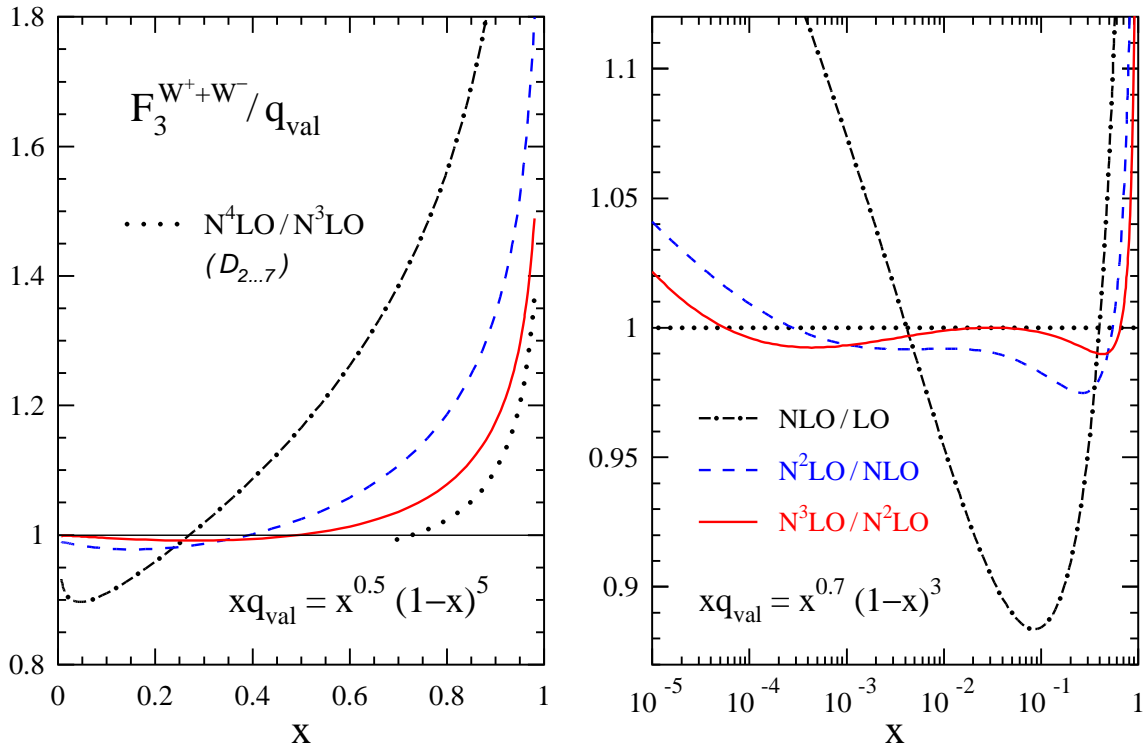,width=15.8cm}\quad}
\vspace{-2mm}
\caption{ \label{pic:cexpcnv3}
 As the right parts of Figs.~\ref{pic:cexpcnv1} and \ref{pic:cexpcnv2},
 but for different choices of the quark distribution~$q_{\rm val\,}$.$\!$} 
\vspace{-1mm}
\end{figure}

Specific numbers as given in the last two paragraphs depend on the quark
distribution. This is illustrated in Fig.~\ref{pic:cexpcnv3}, where the
Reggeon$\,\times\,$counting-rule ansatz (\ref{shape}), cf.~Ref.~\cite
{Reya:1981zk}, for $xq_{\rm val}$ is modified at large $x$ (left) or small $x$
(right). Suppressing $q_{\rm val}$ in either region leads to larger corrections
mainly in the same region, hence Fig.~\ref{pic:cexpcnv3} focuses on the
large-$x$ region in the former case, and small values of $x$ in the latter.
Again we show the results for $\as = 0.2$ and four flavours.

Suppressing $xq_{\rm val}$ by two powers of $(1-x)$ at large-$x$ leads to a
considerable widening of the region of large soft-gluon corrections, with the
NLO, NNLO and N$^3$LO 5\% $x$-values listed above Fig.~\ref{pic:cexpcnv1}
all reduced by about 0.1. A reduction of the small $x$-quark distribution by a
factor $x^{\,0.2}$, i.e., a factor of 10 at $x = 10^{-5}$ with respect to
Eq.~(\ref{shape}) has a rather dramatic effect, especially at NLO, as shown in
the right part of  Fig.~\ref{pic:cexpcnv3}. However, while the relative NLO
correction is as large as 24\% at $x = 10^{-5}$ in this case, the higher-order
corrections remain small, with the N$^3$LO contribution exceeding 1\% only at
$x \leq 2\cdot 10^{-5}$. Hence, even under these conditions, the perturbative
expansion to N$^3$LO proves sufficient for (more than) all practically
relevant situations.

Like the large-$x$ rise in the right part of the figure, the small-$x$ rise 
on the left appears to move closer to the end-point as the order in $\as$ is 
increased. Unlike at large $x$, though, there is no way of estimating the 
fourth-order contributions at values $x \lsim 10^{-4}$, as there was no way
predicting the size of the $\dabc2$ three-loop correction from lower-order 
information. Also fixed moments cannot provide constraints at such values of
$x$, therefore the present (fortunately satisfactory) status will remain 
unless$/$until someone (else) calculates the exact fourth-order coefficient 
function.
%
%
\vspace{-1mm}
\setcounter{equation}{0}
\section{Summary and outlook}
\label{summary}
%
%
We have extended our previous computations \cite{Moch:2004xu,Vermaseren:2005qc} 
of exact third-order coefficient functions in inclusive deep-inelastic 
scattering to the charged-current structure function $F_3^{\,W^+\! + W^-}$. 
Hence the next-to-next-to-next-to-leading order coefficient functions are now 
known, in massless perturbative QCD, for all structure functions for which 
precision measurements have been performed in fixed-target DIS and$/$or at 
HERA, enabling improved analyses of such data at $x > 0.01$.

Also the present calculation has been performed in Mellin-$N$ space, obtaining
an analytic formula in terms of harmonic sums \cite{Vermaseren:1998uu} for 
all odd-$N$ moments (as in Refs.~\cite{Bardeen:1978yd} and \cite{Moch:1999eb}
at first and second order) using the optical theorem and a dispersion relation
in the Bjorken variable $x$. From this result, which we will make available 
but did not write down in this article for brevity, we have reconstructed the 
equally lengthy exact $x$-dependence in terms of harmonic polylogarithms 
\cite{Remiddi:1999ew} presented in Eq.~(A.3). A compact and accurate 
parametrization is provided by Eq.~(\ref{c3ns3}).

The singular large-$x$ terms, $(1-x)_+^{-1}\,\ln^{\,k}(1-x)$ and $\delta(1-x)$,
are the same for $F_3^{\,W^+\!\pm W^-}$ and $F_2^{\,W^+\!\pm W^-}$. The latter
coefficient function, in turn, is identical to that for $F_2^{\,\rm e.m.}$ of 
Ref.~\cite{Vermaseren:2005qc} up to the obvious absence of the $fl_{11}$ 
diagram class which contributes to the coefficient of $\delta(1-x)$ at 
three-loop in the photon- and $Z$-exchange cases. Even at the largest values of
$x$ practically relevant at large scales $\Qs$ and invariant masses $W^{\,2\!}$,
the observable-specific non-leading contributions to the coefficient functions
are non-negligible. We have presented, in particular for use in theoretical 
studies of subleading terms, explicit expressions of the $\ln^{\,k}(1-x)$ 
contributions to both $\Ftwo$ and $\F3$ which turn out to be related via the 
corresponding terms for $\FL$ already presented in 
Ref.~\cite{Vermaseren:2005qc}.

At small-$x$ the N$^{\,l}$LO non-singlet coefficient functions include 
potentially large logarithmic terms up to $\as^{\,l}\, \ln^{\,2l-1}x$. 
For $F_2^{\,\rm e.m.}$ we found that the prefactors of the third-order terms 
are such that a small-$x$ rise only occurs at irrelevantly low values of 
$x$~\cite{Vermaseren:2005qc}. The results for the vector$\,$--$\,$axial-vector 
interference quantity $F_3^{\,W^+\! + W^-}$ would be similar, were it not for 
the $\dabc2$ $\fl02$ diagram class (absent in $\Ftwo$ due to Furry's theorem) 
which occurs at the third order for the first time. These diagrams dominate the
small-$x$ limit and lead to a rise of the coefficient function at $x \lsim 
10^{-3}$ before and $x \lsim 10^{-4}$ after the convolution with the valence 
quark distribution. Nevertheless, the corrections remain very small at 
$x$-values accessible to colliders and unproblematic down even to the very low 
values of $x$ of interest in DIS of ultra-high energy cosmic neutrinos at 
$\Qs \approx \,M_W^{\,2}$.

Progress beyond our present three-loop accuracy for $F_3^{\,W^+\!+W^-}\!$ would
be possible at large-$x$ if the fixed-$N$ results of Ref.~\cite{Retey:2000nq} 
could be extended to the fourth order, by combining those results with the 
substantial constraints from the threshold resummation \cite{Moch:2005ba} into 
approximations analogous to those of Ref.~\cite{vanNeerven:2001pe}. Given that 
a first four-loop moment has been computed already, the splitting function for 
quark combinations such as $u+\bar{u}-(d+\bar{d})$ at $N=2$ for three flavours 
\cite{Baikov:2006ai}, this seems a not entirely unrealistic perspective.
 
On the other hand, any attempt of reliably inferring the small-$x$ behaviour of 
$F_3^{\,W^+\!+W^-}\!$ via a resummation of lower-order information appears 
doomed to failure by the hierarchy of the small-$x$ coefficients, recall 
Eq.~(\ref {c33L0num}), and the possible occurrence of dominant new colour 
factors such as $\dabc2$ in the present three-loop case, see also Ref.~\cite
{Moch:2004pa}. Hence the rather forbidding extension of the present all-$N$ 
computation to the fourth order would be required for progress at small $x$. 
Fortunately the size of the three-loop corrections does not call urgently for 
such a calculation.

{\sc Form} files of our results in both $N$-space and $x$-space, and {\sc 
Fortran} subroutines of the exact and approximate coefficient functions can be 
obtained from the preprint server $\,${\tt http:/}$\!\!${\tt /arXiv.org} by 
downloading the source of this article. Furthermore they are available from us 
upon request.
%
%
\subsection*{Acknowledgments}
S.M. acknowledges support by the Helmholtz Gemeinschaft under contract 
VH-NG-105 and in part by the Deutsche Forschungsgemeinschaft in 
Sonderforschungs\-be\-reich/Transregio~9. 
The work of J.V. has been part of the research program of the Dutch Foundation 
for Fundamental Research of Matter (FOM). The research of A.V. has been 
supported by the UK Science \& Technology Facilities Council (STFC) under grant
numbers PP/E007414/1 and ST/G00062X/1.
%
%
%
\setlength{\baselineskip}{0.54cm}
\setcounter{equation}{0}
\renewcommand{\theequation}{A.\arabic{equation}}
\section*{Appendix A: The exact $x$-space results}
%
%
Here we present the exact expressions for the coefficient functions up to the 
third order in terms of harmonic polylogarithms 
${\rm H}_{\,m_1^{},\ldots,\,m_w}(x)$, $m_j = 0,\pm 1$~\cite{Remiddi:1999ew}. 
Functions up to weight (number of indices) $\,2n-1$ contribute at order 
$\as^{\,n}$.  For `brevity' we suppress the argument $x$ and define 
\beq
\label{app:pqq}
  \pqq(x) \; =\;  2\, (1 - x)^{-1} - 1 - x \; .
\eeq
All divergences for $x \to 1 $ in Eq.~(\ref{app:pqq}) and below are to be read
as $+$-distributions, see Eqs.~(\ref{plus}) and (\ref{Dkconv}) above. In this 
notation the one-loop coefficient function (\ref{c3ns1}) for $\F3$ can be 
written as 
\begin{eqnarray}
\label{eq:c31x}
c_{3,\pm}^{\,(1)}(x) & \! = \! &  
         \colour4colour{\cf}  \* \biggl( \,
            {1 \over 2} \* (5 + x)
          - {1 \over 2} \* \pqq(x) \* (3 + 4 \* \H(0)
          + 4 \* \H(1))
          - \delta(1 - x) \* (9 + 4 \* \z2)
          \biggr)
\:\: .
\end{eqnarray}

\noindent
The exact two-loop result corresponding to Eq.~(\ref{c3ns2}) reads
\begin{eqnarray}
\label{eq:c32x}
&& \!\! c_{3,-}^{\,(2)}(x) \:\: = \:\: 
         \colour4colour{\cfs}  \*  \biggl(
          - 4 \* \pqq( - x) \* (7 \* \z3 - 8 \* \H(-1) \* \z2 - 2 \* \H(0) 
          + 2 \* \H(0) \* \z2 - 8 \* \Hhh(-1, -1,0) + 10 \* \Hhh(-1,0,0) 
  \nonumber\\[-1mm]&& \mbox{}
         + 4 \* \Hhh(-1,0,1) + 6 \* \Hhh(0,-1,0) - 3 \* \Hhh(0,0,0)
          - 2 \* \Hhh(0,0,1))
          + {1 \over 4} \* \pqq(x) \* (51 + 128 \* \z3 + 48 \* \z2 + 122 \*
          \H(0) + 96 \* \H(0) \* \z2 
  \nonumber\\&& \mbox{}
          + 54 \* \H(1) + 32 \* \H(1) \* \z2 - 12 \* \Hh(0,0) - 48 \* \Hh(0,1) 
          - 72 \* \Hh(1,0) - 72 \* \Hh(1,1)
          + 96 \* \Hhh(0,-1,0) - 32 \* \Hhh(0,0,0) - 96 \* \Hhh(0,0,1) 
  \nonumber\\&& \mbox{}
          - 96 \* \Hhh(0,1,0) - 
         112 \* \Hhh(0,1,1) - 48 \* \Hhh(1,0,0) - 96 \* \Hhh(1,0,1) 
          - 128 \* \Hhh(1,1,0) - 96 \* \Hhh(1,1,1))
          + 4 \* (3 - x) \* \z3
  \nonumber\\&& \mbox{}
          - 8 \* (1 + 2 \* x^{-1} + x + 2 \* x^2) \* \Hh(-1,0)
          + 8 \* (1 - x) \* (\H(1) \* \z2 - \Hhh(1,0,0))
          - 2 \* (1 + x) \* (4 \* \H(-1) \* \z2 + 4 \* \H(0) \* \z2 
  \nonumber\\&& \mbox{}
          - 3 \* \Hh(1,0) 
          + 8 \* \Hhh(-1,-1,0) - 4 \* \Hhh(-1,0,0) - 5 \* \Hhh(0,0,0) 
          - 4 \* \Hhh(0,0,1) - 2 \* \Hhh(0,1,0) - 2 \* \Hhh(0,1,1))
          + 2 \* (1 + 5 \* x) \* \Hh(1,1)
  \nonumber\\&& \mbox{}
          - {1 \over 2} \* (3 + 79 \* x) \* \H(0)
          - 8 \* (4 + 5 \* x + 2 \* x^2) \* \z2
          + {1 \over 2} \* (39 - 17 \* x) \* \H(1)
          + (45 + 49 \* x + 16 \* x^2) \* \Hh(0,0)
  \nonumber\\&& \mbox{}
          + 8 \* (3 + 5 \* x) \* \Hh(0,1)
          - {1 \over 4} \* (233 - 131 \* x)
          + 16 \* \Hhh(0,-1,0) \* x
          + \biggl\{ {331 \over 8} - 78 \* \z3 + 69 \* \z2 + 6 \* \z2^2 
          \biggr\} \* \delta(1-x)
          \biggr)
  \nonumber\\&& \mbox{}
       + \colour4colour{\cf \* \ca}  \*  \biggl(
          - 8 \* \z3
          + 2 \* \pqq( - x) \* (7 \* \z3 - 8 \* \H(-1) \* \z2 - 2 \* \H(0) 
          + 2 \* \H(0) \* \z2 - 8 \* \Hhh(-1,-1,0) + 10 \* \Hhh(-1,0,0) 
  \nonumber\\[-1mm]&& \mbox{}
         + 4 \* \Hhh(-1,0,1) 
         + 6 \* \Hhh(0,-1,0) - 3 \* \Hhh(0,0,0)
          - 2 \* \Hhh(0,0,1))
          - {1 \over 108} \* \pqq(x) \* (3155 - 216 \* \z3 - 1584 \* \z2 
          + 4302 \* \H(0) 
  \nonumber\\&& \mbox{}
          - 432 \* \H(0) \* \z2
          + 2202 \* \H(1) - 1296 \* \H(1) \* \z2 + 1980 \* \Hh(0,0) 
          + 1584 \* \Hh(0,1) + 1296 \* \Hhh(0,-1,0) + 648 \* \Hhh(0,0,0) 
  \nonumber\\&& \mbox{}
         + 792 \* \Hh(1,0) + 792 \* \Hh(1,1) 
         + 432 \* \Hhh(0,0,1)
         + 864 \* \Hhh(1,0,0) + 432 \* \Hhh(1,0,1) - 432 \* \Hhh(1,1,0))
         - {9 \over 2} \* (3 - 5 \* x) \* \H(1)
  \nonumber\\&& \mbox{}
          + 4 \* (1 + 2 \* x^{-1} + x + 2 \* x^2) \* \Hh(-1,0)
          - 4 \* (1 - x) \* (\H(1) \* \z2 - \Hhh(1,0,0))
          - 4 \* (4 + 3 \* x + 2 \* x^2) \* \Hh(0,0)
  \nonumber\\&& \mbox{}
          + 4 \* (1 + x) \* (\H(-1) \* \z2 - 2 \* \Hh(0,1) 
          + 2 \* \Hhh(-1,-1,0) - \Hhh(-1,0,0))
          + 4 \* (3 + 2 \* x + 2 \* x^2) \* \z2
          - {1 \over 6} \* (29 - 271 \* x) \* \H(0)
  \nonumber\\&& \mbox{}
          + {5 \over 36} \* (91 + 167 \* x)
          - 8 \* \Hhh(0,-1,0) \* x
          + \biggr\{ - {5465 \over 72} + {140 \over 3} \* \z3 - {251 \over 3}
          \* \z2 
          + {71 \over 5} \* \z2^2\biggl\} \* \delta(1-x)
          \biggr)
  \nonumber\\&& \mbox{}
       + \colour4colour{\cf \*  \nf}  \*  \biggl(
            {1 \over 54} \* \pqq(x) \* (247 - 144 \* \z2 + 342 \* \H(0) 
          + 174 \* \H(1) + 180 \* \Hh(0,0)
          + 144 \* \Hh(0,1) + 72 \* \Hh(1,0) + 72 \* \Hh(1,1))
  \nonumber\\&& \mbox{}
          + {1 \over 3} \* (1 - 11 \* x) \* \H(0)
          + (1 - 3 \* x) \* \H(1)
          + {1 \over 18} \* (5 - 119 \* x)
          + \biggr\{ {457 \over 36} + {38 \over 3} \* \z2 
          + {4 \over 3} \* \z3 \biggr\} \* \delta(1-x)
          \biggr)
\:\: .
\end{eqnarray}

\noindent
Finally full three-loop coefficient function for $F_3^{\,W^+\! + W^-}$ 
approximated by Eq.~(\ref{c3ns3}) is given by
{\small 
\setlength{\baselineskip}{0.58cm}

}

\setlength{\baselineskip}{0.54cm}
\noindent
Eq.~(\ref{eq:c33x}) includes, with the colour factor $\cf (\ca-2\,\cf)^2$, the
sum of the inverse Mellin transforms of the functions $\gfunct1(N)$ and
$\gfunct2(N)$ in Eqs.~(\ref{g12N}). These functions are given by
\begin{eqnarray}
  \label{eq:g1x}
&& \!\!\!\!\gfunct1(x) \: =\: 
       - 2 \* (1 - (1-x)^{-1}) \* (
            4 \* \H(-1) \* \z2 
          - 2 \* \Hhh(-1,0,0) 
          - 4 \* \Hhh(-1,0,1) 
          + \Hhh(0,0,0)
          + 2 \*  \Hhh(0,0,1)
          - 3 \* \H(0) \* \z2 
          - 3 \* \z3
          )
  \nonumber\\&& \mbox{}
       + 2 \* ((1-x)^{-2} - (1-x)^{-1}) \* \biggl(
           4 \* \Hh(0,-1) \* \z2 
         - 2 \* \Hhhh(0,-1,0,0) 
         - 4 \*  \Hhhh(0,-1,0,1) 
         + 2 \* \Hhhh(0,0,0,1)
         - 3 \* \Hh(0,0) \* \z2
  \nonumber\\&& \mbox{}
         - 3 \* \H(0) \* \z3 
         + \Hhhh(0,0,0,0)
         - {2 \over 5} \* \zss
         \biggr)
       + 2 \* (1 - (1+x)^{-1}) \* ( 2 \* \Hhh(0,-1,0) 
       - \Hhh(0,0,0) + \H(0) \* \z2 + 2 \* \z3 )
  \nonumber\\&& \mbox{}
       - ((1+x)^{-2} - (1+x)^{-1}) \* \biggl(
           4 \* \Hhhh(0,0,-1,0)
         + 2 \* \Hh(0,0) \* \z2
         + 4 \* \H(0) \* \z3 
         - 2 \* \Hhhh(0,0,0,0)
         + {21 \over 5} \* \zss
         \biggr) 
\:\: ,
\\[2mm]
  \label{eq:g2x}
&& \!\!\!\!\gfunct2(x) \: = \: 
       - 2 \* (2 \* (1-x)^{-3} - 3 \* (1-x)^{-2} + (1-x)^{-1}) \* \biggl(
           4 \* \Hh(0,-1) \* \z2 
         + \Hhhh(0,0,0,0) 
         - 2 \* \Hhhh(0,-1,0,0) 
  \nonumber\\[-1mm]&& \mbox{}
         - 4 \* \Hhhh(0,-1,0,1) 
         + 2 \* \Hhhh(0,0,0,1)
         - 3 \* \Hh(0,0) \* \z2
         - 3 \* \H(0) \* \z3
         - {2 \over 5} \* \zss
         \biggr)
       - 2 \* \z2 \* (1 - (1-x)^{-1})
  \nonumber\\&& \mbox{}
       - 4 \* ((1-x)^{-2} - (1-x)^{-1}) \* (
            4 \* \H(-1) \* \z2 
          - 2 \* \Hhh(-1,0,0) 
          - 4 \* \Hhh(-1,0,1) 
          + 2 \* \Hhh(0,0,1) 
          + \Hhh(0,0,0)
          - 3 \* \H(0) \* \z2
          - 3 \* \z3
          )
  \nonumber\\&& \mbox{}
       + (2 \* (1+x)^{-3} - 3 \* (1+x)^{-2} + (1+x)^{-1}) \* \biggl(
           4 \* \Hhhh(0,0,-1,0)
         - 2 \* \Hhhh(0,0,0,0) 
         + 2 \* \Hh(0,0) \* \z2
         + 4 \* \H(0) \* \z3
         + {21 \over 5} \* \zss
         \biggr) 
  \nonumber\\&& \mbox{}
       + 4 \* ((1+x)^{-2} - (1+x)^{-1}) \* ( 
            2 \* \Hhh(0,-1,0) 
          - \Hhh(0,0,0)
          + \H(0) \* \z2
          + 2 \* \z3
          )
  \nonumber\\[1mm]&& \mbox{}
       - 2 \* (1 - (1+x)^{-1}) \* (2 \* \Hh(-1,0) + 2 \* \Hh(0,1) - \z2)
       + (\z2 - \z3) \* \delta(1-x)
\:\: .
\end{eqnarray}
Note that the $1/(1-x)$ terms in Eqs.~(\ref{eq:g1x}) and (\ref{eq:g2x}) do 
not imply the presence of +-distributions, but large-$x$ cancellations between
the harmonic polylogarithms in the round brackets. These cancellations become
numerically problematic for $1-x \ll 1$, a region unavoidable in moment and 
Mellin convolution integrals, even with the high-accuracy code of Ref.~\cite
{Gehrmann:2001pz}. In this region one can instead use
\bea
\label{eq:g12xto1}
 (\gfunct1 + \gfunct2)(x) &\! = \!& (\z2 - \z3)\, \delta(1-x)
 \:-\: (1-x)\, \bigg\{ \ln (1-x) + {\z2 \over 2} - \z3 - {3 \over 4} \bigg\}
\nn \\ & & 
 \:-\: {1 \over 2}\: (1-x)^2 \:+\: {\cal O}[(1-x)^3] \:\: .
\eea
%
%
%
\vspace*{-3mm}
\renewcommand{\theequation}{B.\arabic{equation}}
\setcounter{equation}{0}
\section*{Appendix B: Subleading large-$x$ contributions to $F_{\:\!2,\rm ns}$}
In this final appendix we consider the subleading (integrable) large-$x$
logarithms $\ln^{\,k}(1-x)$ for the quark contributions to $\Ftwo$ and their 
relation to the coefficients for $\F3$ given in Eqs.~(\ref{c32L13}) -- 
(\ref{c33L11}) above. The corresponding three-loop results for $\FL$ (where 
these terms are the leading contributions) have been given in Eq.~(4.31) - 
(4.34) of Ref.~\cite{Vermaseren:2005qc}. 
At two loops we have 
\bea
\label{c22L13}
  c_{\,2,\rm q\,}^{\,(2)} \Big|_{\,L_1^3} \!\!\! & = \! &
          - \: 8\: \* \cfs
\\[0.5mm]
\label{c22L12}
  c_{\,2,\rm q\,}^{\,(2)} \Big|_{\,L_1^2} \!\!\! & = \! &
            {22 \over 3}\: \* \ca \* \cf
       \: + \: 60\: \* \cfs
       \: - \: {4 \over 3}\: \* \cf\, \* \nf 
\\[0.5mm]
\label{c22L11}
  c_{\,2,\rm q\,}^{\,(2)} \Big|_{\,L_1} \!\!\! & = \! &
       - \: \ca \* \cf \*  \:\Bigg[
            {916 \over 9}
          - 24\, \* \z2
         \Bigg] 
     \: + \: 20 \* \cfs 
     \: + \: {148 \over 9}\: \* \cf\, \* \nf  
\:\: .
\eea
 
The corresponding contributions to the third-order coefficient function are 
given by
\bea
\label{c23L15}
  c_{\,2,\rm q\,}^{\,(3)} \Big|_{\,L_1^5} \!\!\! & = \! &
          - \: 8\: \* \cft
\\[0.5mm]
\label{c23L14}
  c_{\,2,\rm q\,}^{\,(3)} \Big|_{\,L_1^4} \!\!\! & = \! &
            {220 \over 9} \: \* \ca \* \cfs
       \: + \: 92\: \* \cft
       \: - \: {40 \over 9} \: \* \cfs\, \* \nf
\\[1mm]
\label{c23L13}
  c_{\,2,\rm q\,}^{\,(3)} \Big|_{\,L_1^3} \!\!\! & = \! &
       - \: {484 \over 27}\: \* \cas \* \cf
       - \: \ca \* \cfs \*  \:\Bigg[
            {10976 \over 27}
          - 64\, \* \z2
         \Bigg]
     \: - \: \cft  \* \: [
            38
          - 32\, \* \z2
          ]
\nn \\[1mm] & & \mbox{\hspn}
     \: + \: {176 \over 27}\: \* \ca \* \cf\, \* \nf  
     \: + \: {1832 \over 27}\: \* \cfs\, \* \nf
     \: - \: {16 \over 27}\: \* \cf\, \* \nfs
\\[2mm]
\label{c23L12}
  c_{\,2,\rm q\,}^{\,(3)} \Big|_{\,L_1^2} \!\!\! & = \! &
          \cas \* \cf \* \:\Bigg[
            {11408 \over 27}
          - {266 \over 3}\: \* \z2
          - 32\, \* \z3
         \Bigg]
     \: + \: \ca \* \cfs \* \:\Bigg[ 
            {11501 \over 9}
          - 292\, \* \z2
          - 160\, \* \z3
          \Bigg]
\nn\\[1mm] & & \mbox{\hspn}
     - \: \cft \* \:\Bigg[
            {1199 \over 3}
          + 688\, \* \z2
          + 48\, \* \z3
          \Bigg]
     \: - \: \ca \* \cf\, \* \nf \* \:\Bigg[ 
            {3694 \over 27}
          - {64 \over 3}\: \* \z2
          \Bigg]
\nn\\[1mm] & & \mbox{\hspn}
     - \: \cfs\, \* \nf \* \:\Bigg[
            {2006 \over 9}
          - {16 \over 3}\, \* \z2
          \Bigg]
     \: + \: {296 \over 27}\: \* \cf\, \* \nfs
\\[2mm]
\label{c23L11}
  c_{\,2,\rm q\,}^{\,(3)} \Big|_{\,L_1} \!\!\! & = \! &
        - \: \cas \* \cf \* \:\Bigg[
            {215866 \over 81}
          - 824\: \* \z2
          - {1696 \over 3}\, \* \z3
          + {304 \over 5}\: \* \zss
         \Bigg]
     \: + \: \ca \* \cfs \* \:\Bigg[
            {126559 \over 162}
\nn\\[1mm] & & 
          + 872\: \* \z2
          + 792\: \* \z3
          - {1916 \over 5}\: \* \zss
          \Bigg]
     \: + \: \cft  \*  \:\Bigg[
            {157 \over 6}
          + {1268 \over 3}\: \* \z2
\nn\\[1mm] & & 
          - 272\, \* \z3
          + 488\: \* \zss
          \Bigg]
     \: + \: \ca \* \cf\, \* \nf \* \:\Bigg[
            {64580 \over 81}
          - {1292 \over 9}\, \* \z2
          - {304 \over 3}\, \* \z3
          \Bigg]
\nn\\[1mm] & & \mbox{\hspn}
     \: - \: \cfs\, \* \nf \* \:\Bigg[
            {4445 \over 81}
          + 208\: \* \z2
          - {208 \over 3}\: \* \z3
          \Bigg]
     \: - \: \cf\, \* \nfs \* \:\Bigg[
            {4432 \over 81}
          - {32 \over 9}\: \* \z2
          \Bigg]
\:\: .
\eea
 
A comparison of the large-$x$ limit of the complete charged-current quark 
coefficient functions $C_{1,\:2,\:3,\:L\,}$, defined analogous to 
Eq.~(\ref{C3min}) above, now reveals that
\beq
\label{eq:C13rel}
 C_1(x,\as) \:\; \equiv \:\; C_2(x,\as) - C_L(x,\as) 
            \:\; = \:\; C_3(x,\as) \: + \: {\cal O}(1-x) 
\eeq
to the third order in $\as$. I.e., not only are the above log-enhanced terms of 
$C_1$ and $C_3$ related, but also the constants for $x \ra 1$, leaving the 
difference vanish as $1/N^{\,2}$ in moment space. To the best of our knowledge,
Eq.~(\ref{eq:C13rel}) has not been noted so far in the literature even on the
level of the  two-loop coefficient functions of 
Refs.~\cite{vanNeerven:1991nn,Zijlstra:1991qc,Zijlstra:1992kj}. In this context
it appears worthwhile to recall that the non-singlet coefficient function for 
the polarized structure function $g_1^{}$ [no relation to the quantity in 
Eq.~(\ref{eq:g1x})] is identical to $C_3$ up to the $\dabc2$ term irrelevant at
large $x$, and that an analogous $1/N^{\,2}$ suppression of the helicity flip 
for $x \ra 1$ is present in polarized splitting functions, see 
Ref.~\cite{Vogt:2008xx} and references therein.
%

{\footnotesize
\setlength{\baselineskip}{0.5cm}

}

\end{document}